\documentclass[a4paper,fleqn,usenatbib,useAMS,english]{mnras}
\usepackage[T1]{fontenc}
\usepackage{ae,aecompl}
\usepackage[usenames,dvipsnames,svgnames,table]{xcolor}
\usepackage{hyperref}
\usepackage{amssymb}
\usepackage{amsmath}
\usepackage{babel}
\usepackage{graphicx}
\usepackage{multicol}
\usepackage{bm}
\usepackage{pdflscape}
\usepackage{rotating}
\usepackage{longtable}
\usepackage[ruled,vlined]{algorithm2e}

\makeatletter

\newcommand{\sn}[2]{\ensuremath{#1 \times 10^{#2}}} 
\newcommand{\qq}{$\arcsec$} 
\newcommand{\dsq}{deg$^2$} 

\DeclareMathAlphabet\mathbfcal{OMS}{cmsy}{b}{n} 
\newcommand{\rom}[2]{\ensuremath{#1_{\rm #2}}} 
\newcommand{\Normal}[3]{\ensuremath{\mathcal{N}\left({#1} \,\middle|\, {#2}, {#3} \right)}} 

\newcommand{\spz}{spec-$z$} 
\newcommand{\prz}{prism-$z$} 
\newcommand{\grz}{grism-$z$} 
\newcommand{\gprz}{g/prism-$z$} 
\newcommand{\phz}{photo-$z$} 

\newcommand{\hsc}{HSC-SSP}
\newcommand{\hscwide}{\texttt{Wide}}
\newcommand{\hscdeep}{\texttt{Deep}}
\newcommand{\hscudeep}{\texttt{UltraDeep}} 

\title[Validating GG Lensing for HSC]{Galaxy-Galaxy Lensing in HSC: 
Validation Tests and the Impact of Heterogeneous Spectroscopic Training Sets}

\author[Speagle et al.]{
    Joshua S. Speagle$^{1}$\thanks{NSF Graduate Research Fellow}\thanks{E-mail: jspeagle@cfa.harvard.edu},
	Alexie Leauthaud$^{2}$,
    Song Huang$^{2}$, \and
    Christopher P. Bradshaw$^{2}$,
    Felipe Ardila$^{2}$,
    Peter L. Capak$^{3}$, \and
    Daniel J. Eisenstein$^{1}$,
	Daniel C. Masters$^{4}$,
    Rachel Mandelbaum$^{5}$, \and
    Surhud More$^{6,7}$,
    Melanie Simet$^{8,9}$,
    and Crist\'obal Sif\'on$^{10}$
	\\
    \\
	$^{1}$Harvard University, 60 Garden St., MS 46, Cambridge, MA 02138, USA\\
	$^{2}$University of California Santa Cruz, 1156 High St., Santa Cruz, CA 95064, USA\\
 	$^{3}$Spitzer Science Center, California Institute of Technology, Pasadena, CA 91125, USA\\
    $^{4}$Infrared Processing and Analysis Center, California Institute of Technology, 
    Pasadena, CA 91125, USA\\
    $^{5}$McWilliams Center for Cosmology, Department of Physics, Carnegie Mellon University, 
    Pittsburgh, PA 15213, USA\\
    $^{6}$Kavli Institute for the Physics and Mathematics of the Universe (Kavli IPMU, WPI), University
    of Tokyo, Chiba 277-8582, Japan\\
    $^{7}$The Inter-University Center for Astronomy and Astrophysics, Post bag 4, Ganeshkhind,
    Pune, 411007, India\\
    $^{8}$University of California, Riverside, 900 University Avenue, Riverside, CA 92521, USA\\
    $^{9}$Jet Propulsion Laboratory, California Institute of Technology, Pasadena, CA 91109, USA\\
    $^{10}$Department of Astrophysical Sciences, Princeton University, 
    Peyton Hall, 4 Ivy Ln, Princeton, NJ 08544, USA
}

\date{Accepted XXX. Received YYY; in original form ZZZ}

\pubyear{2018}

\begin{document}
\label{firstpage}
\pagerange{\pageref{firstpage}--\pageref{lastpage}}
\maketitle
	
\begin{abstract}
Although photometric redshifts (\phz's) are crucial ingredients for current and upcoming 
large-scale surveys, the high-quality spectroscopic redshifts currently available to train, validate, and test 
them are substantially non-representative in both magnitude and color. We investigate the nature and structure 
of this bias by tracking how objects from a heterogeneous training sample contribute to 
{\phz} predictions as a function of magnitude and color, and illustrate that the underlying 
redshift distribution at fixed color can evolve strongly as a function of magnitude. 
We then test the robustness of the galaxy-galaxy lensing signal in 120 deg$^2$ of {\hsc} DR1 data
to spectroscopic completeness and {\phz} biases, and find that their impacts are 
sub-dominant to current statistical uncertainties. Our methodology provides a framework
to investigate how spectroscopic incompleteness can impact {\phz}-based weak lensing predictions in 
future surveys such as LSST and WFIRST.
\end{abstract}




\section{Introduction}
\label{sec:intro}


Between the surface of last scattering ($z \sim 1100$) and the present day ($z=0$), the paths of all 
observed photons have been gravitationally influenced by the intervening ``cosmic web'' of matter. This 
gravitational lensing, and particularly weak lensing, is sensitive to the growth of structure 
and expansion history of the Universe and serves as a key probe of cosmology 
\citep[e.g., see review by][]{mandelbaum18}. 
In addition, weak lensing serves as an effective complementary technique to other cosmological probes 
(e.g., the Cosmic Microwave Background or Type Ia supernovae) by helping to break degeneracies
between cosmological parameters and providing constraints on the growth of large scale structure 
\citep[e.g.,][for some recent cosmic shear results]{des17,hildebrandt+18,hikage+18}.

The determination of accurate photometric redshifts (\phz's) is a key challenge for deep 
lensing surveys. While shallow surveys ($i<24$) can obtain spectroscopic follow-up for representative samples,
deeper surveys face tougher challenges. In this paper, we focus on the challenges of deriving \phz's for
the first year source catalog of the HSC\footnote{The Hyper Suprime-Cam (HSC) 
Subaru Strategic Program (SSP) Survey \citep{aihara+18}. See \url{hsc.mtk.nao.ac.jp/ssp}.} survey
which reaches an $i$-band depth of $\sim 26$ AB magnitudes. As a precursor to 
LSST\footnote{The Large Synoptic Survey Telescope \citep{ivezic+08}. See \url{lsst.org}.}, 
the {\hsc} survey is a crucial testing ground
for {\phz} methods that will be applied for future precision cosmology analyses.

At the depths probed by HSC, there is a lack of adequate representative spectroscopic redshifts
(\spz's) available for training, validating, and testing {\phz} methods \citep{masters+15,tanaka+18}. 
As a result, the HSC {\phz} team instead supplements \spz's taken from a variety of public surveys with 
grism/prism-based redshifts (\gprz's) along with \phz's derived from deep, many-band photometry
when training various {\phz} algorithms and validating their performance. 
These unavoidable choices lead to a heterogeneous training set spanning a wide range of 
possibly redshift ``quality''.

Although mixing \spz's and high-quality alternatives will likely occur in 
future surveys, the impact of using such a
heterogeneous mixture on weak lensing has not yet been extensively explored. 
While the lack of high-quality \spz's
in regions of color and magnitude space makes it difficult to validate {\phz} performance in those regions
independently of the assumptions used to generate them, supplementing \spz's in these regions with
other methods that rely more heavily on these assumptions \citep[see, e.g.,][]{bezanson+16} 
will not alleviate this problem. This means that performance in these regions remains a
``known unknown'' that is difficult to directly validate. This problem is particularly acute for 
future cosmology surveys hoping to derive unbiased \phz's at the sub-percent level to the majority 
of their faint photometric samples.

Currently, there are several attempts in the literature to try to resolve this issue. These take two broad
approaches. The first is an attempt to efficiently collect \spz's to ``fill in'' regions of color space
that currently do not have available data. The largest systematic 
approach is the C3R2 survey \citep{masters+17}, which has so far
collected $\sim \, 1300$ high-quality spectra in under-populated regions of color-space.
The second is to assume that we can use cross-correlations of ensembles of galaxies that span 
the relevant redshift range, regardless of their color and/or magnitude, 
to characterize {\phz} accuracy for a population 
of galaxies. This has proven to be promising but is not without challenges
\citep{menard+13,newman+15,hoylerau18}. Importantly, both of these methods assume that we can use 
\textit{ensembles} of galaxies in specific regions of color and/or magnitude space to calibrate 
{\phz} biases and uncertainties.

In this paper, we investigate how the use of heterogeneous training samples affects {\phz} performance
and galaxy-galaxy (gg) lensing analyses \citep[e.g.,][]{kwan2017,leauthaud+17,prat2018} 
using {\hsc} data. In \S\ref{sec:data}, we describe the
photometry, shear, and redshift data used in this paper. In \S\ref{sec:som}, we investigate the 
representativeness of current {\spz} samples and examine the dependence on color and magnitude. 
We find strong evidence for evolution in the 
redshift distribution of galaxies \textit{at fixed color} as a function of magnitude. 
This leads us to develop a new framework, described in \S\ref{sec:photoz}, for computing \phz's from
heterogeneous data that incorporates magnitude dependence and allows us to track how \textit{specific}
training objects contribute to \textit{individual} {\phz} predictions. We investigate the accuracy 
of {\phz} probability density functions (PDFs) computed using this method
in \S\ref{sec:pzchecks}. 

After discussing our {\phz} tests, in \S\ref{sec:ggtheory} we outline the framework used 
for our lensing analysis. In \S\ref{sec:gglens}, we then test whether
our gg lensing measurements are robust to a variety of estimators, quality cuts, and spectroscopic
incompleteness. We conclude in \S\ref{sec:conc}.

We assume a flat $\Lambda$CDM cosmology whenever appropriate
with $\Omega_\Lambda = 0.7$, $\Omega_{\rm m}=0.3$, and $h=0.7$.
All magnitudes in the paper are AB magnitudes \citep{okegunn83}.
For our lensing calculations (see \S\ref{sec:gglens}), we assume physical coordinates to 
compute $\Delta\Sigma$ in 10 logarithmically spaced bins from 0.05\,Mpc to 15\,Mpc.

\section{Data}
\label{sec:data}

\subsection{The HSC Survey}

The Hyper Suprime-Cam Subaru Strategic Program (\hsc) Survey \citep{aihara+18} utilizes the 
Hyper Suprime-Cam prime-focus camera \citep{miyazaki+18,komiyama+18,kawanomoto+18,furusawa+18}
on the  8.2\,m  Subaru telescope at Mauna Kea. 
The survey has a ``wedding cake'' construction, with three different area/depth combinations to optimize
a variety of science goals: 
the {\hscwide} survey will cover 1400\,{\dsq} in $grizy$ to a limiting depth of $i \sim 26$\,mag, 
the {\hscdeep} survey will cover 26\,{\dsq} to $i \sim 27$\,mag, 
and the {\hscudeep} survey will cover 3.5\,{\dsq} to a depth of $i \sim 28$\,{\dsq}. 
This work is based on the \texttt{S16A} {\hsc} internal data release,
which covers 136.9 \,{\dsq} to full {\hscwide} depths in all five bands.
For more information on the {\hsc} survey, please see \citet{aihara+18}. 
For more information on the data processing and pipeline, please see \citet{bosch+18}.

\subsection{The weak lensing source sample}
\label{subsec:source}

Our sample of galaxy sources is selected using the weak lensing cuts
outlined in \citet{mandelbaum+18}.
In brief, these are a series of quality cuts to ensure that composite model 
(i.e. \texttt{CModel}) photometry, point spread functions (PSFs), 
and measured object shapes are reliable.
Observations are restricted to $i < 24.5$\,mag to avoid 
using data with possibly unreliable shape measurements
and to ensure ``reasonable'' {\spz} coverage (although see \S\ref{sec:som}). 
A ``full-depth, full-color'' (FDFC) cut was also imposed to eliminate sources 
that were not observed in all bands to full depth.
Objects near bright stars were removed using the updated
\texttt{Arcturus} bright star masks described in \citet{coupon+18}, as opposed to the original
\texttt{Sirius} masks used in \citet{mandelbaum+18}, as those preserve more galaxies around bright stars.
See \citet{mandelbaum+18} for additional details regarding
the construction and validation of the {\hsc} \texttt{S16A} weak lensing shear catalog.

In addition to these weak lensing cuts, the {\phz}'s used in this work
were only computed for objects with PSF-matched 1.1{\qq} aperture photometry 
available in all five bands. 
This effectively imposes an additional \textit{de facto} cut on the seeing in all five bands. 
A variety of internal tests have found that this does not introduce a meaningful bias on weak 
lensing analyses (More et al., in prep.). 

\subsection{Redshift Training Data}
\label{subsec:train}

\begin{figure*}
\begin{center}
\includegraphics[width=\textwidth]{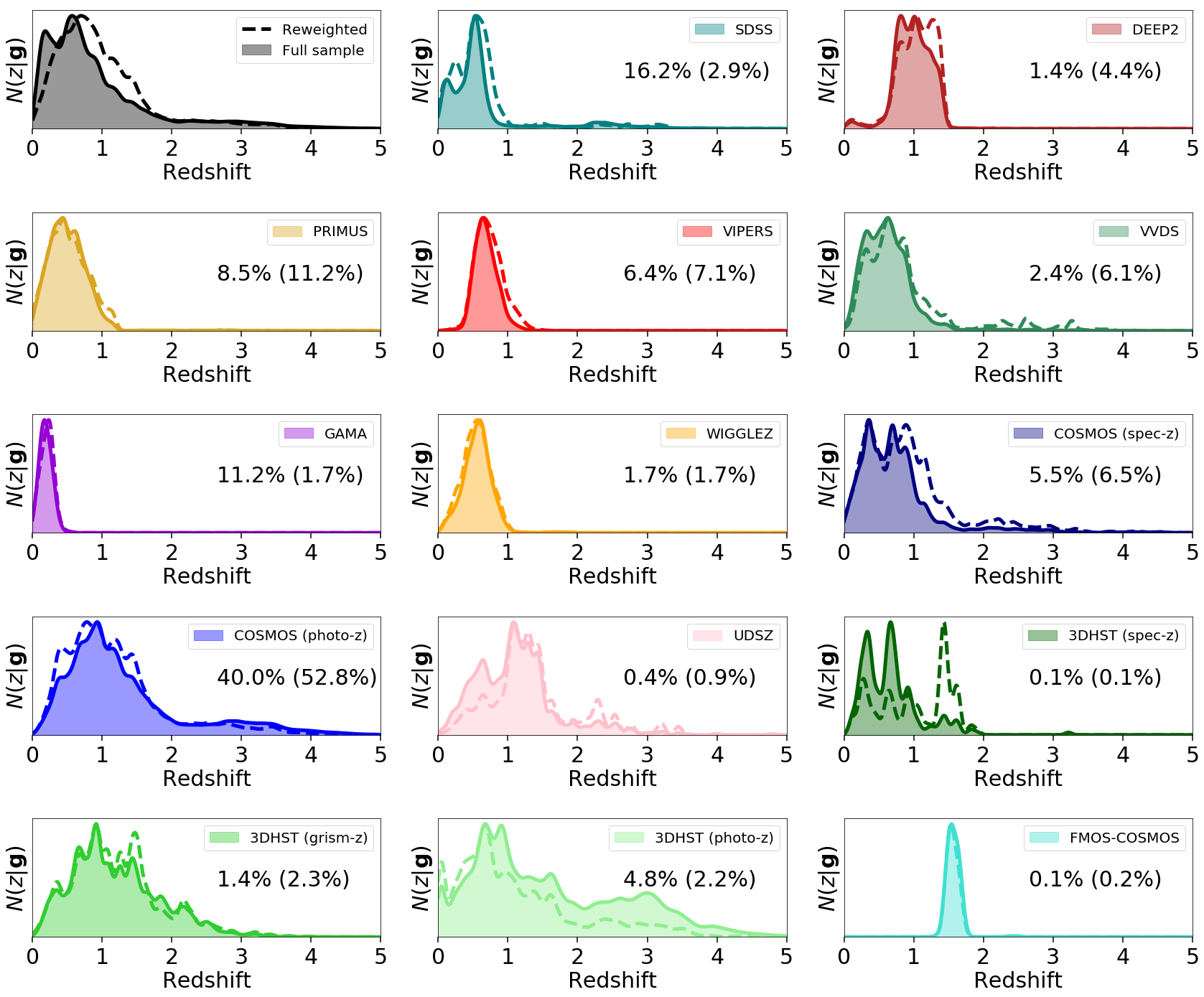}
\end{center}
\caption{The redshift number density $N(z|\mathbf{g})$ of the objects $\mathbf{g}$ included in the {\hsc} 
\texttt{S16A} redshift training set, broken down by survey of origin and source type (if appropriate). 
Solid lines show the original distribution of objects in the training sample, while dashed curves show
the distribution after re-weighting the sample to mimic the color-magnitude distribution of 
the {\hsc} \texttt{S16A} weak lensing photometric sample 
(see \S\ref{subsec:train} and \S\ref{subsec:prior}).
The raw percentage (reweighted percentage) 
of the entire training set (gray, top left) comprised of each subsample is also listed in each figure. 
The color-magnitude weights shift the distribution to higher redshifts and favors deeper surveys
with more diverse galaxy populations (e.g., DEEP2, VVDS) 
over shallower surveys (e.g., GAMA) and those with
more limited galaxy population coverage (e.g., SDSS/BOSS).
\textit{Over 50\% of the objects in the weak lensing sample 
are primarily trained by COSMOS {\phz}}. This
illustrates the paucity of current {\spz} coverage at the depths probed by the {\hsc} survey.
}\label{fig:training}
\end{figure*}

The {\hsc} \texttt{Wide} survey footprint is designed to maximize the overlap with other photometric and 
spectroscopic surveys while keeping the survey geometry simple. This allows {\hsc} to exploit a large number 
of public {\spz}'s when constructing a redshift training set. 
In addition, (\texttt{Ultra})\texttt{Deep} 
data taken in heavily observed fields such as COSMOS \citep{scoville+07} further allows {\hsc} 
to include large numbers of fainter objects observed at higher signal-to-noise (S/N). 
These observations allow for more detailed modeling of the general population observed
in the \texttt{Wide} survey, and are especially helpful for fainter sources.

A detailed description of the training sample can be found in \citet{tanaka+18}. We briefly summarize it
here. The training set contains spectroscopic, grism, and prism redshifts from a variety of overlapping public
surveys including:
\begin{itemize}
\item zCOSMOS DR3 \citep{lilly+09},
\item UDSz \citep{bradshaw+13,mclure+13}, 
\item 3D-HST \citep{skelton+14,momcheva+16}, 
\item FMOS-COSMOS \citep{silverman+15}, 
\item VVDS \citep{lefevre+13}, 
\item VIPERS PDR1 \citep{garilli+14}, 
\item SDSS DR12 \citep{alam+15}, 
\item GAMA DR2 \citep{liske+15}, 
\item WiggleZ DR1 \citep{parkinson+12}, 
\item DEEP2 DR4 \citep{newman+13b}, and
\item PRIMUS DR1 \citep{coil+11,cool+13}. 
\end{itemize}

As each survey has its own flagging scheme to indicate redshift confidence, the
different schemes were homogenized 
and used to select ``secure'' redshifts using the criteria outlined in \citet{tanaka+18}.
In addition to these public surveys, a collection of private COSMOS {\spz}'s
were also included exclusively for {\phz} training (Mara Salvato \& Peter Capak, 
private communication).

In addition to these {\spz}'s, {\grz}'s, and {\prz}'s, the training set was supplemented 
with a set of high-quality, many-band {\phz}'s taken from 3D-HST and COSMOS2015 \citep{laigle+16}
in order to maintain sufficient magnitude and color coverage down to $i \sim 24.5$ (see \S \ref{sec:som}). 
Without these {\phz}'s, the magnitude and color coverage of the training set fails to
adequately span the relevant parameter space of the {\hsc} data used in this analysis. 
The above heterogeneity in the {\spz}'s, {\grz}'s, {\prz}'s, and many-band {\phz}'s 
is one of the motivating reasons
for the analysis presented in this work.

Objects were iteratively matched to this catalog within 1{\qq} at (1) \texttt{UltraDeep}, (2) \texttt{Deep}, 
and (3) \texttt{Wide} {\hsc} depths in order to take advantage of higher-S/N data when available 
while avoiding possible duplicates. Our training data ultimately consists of $\sim$ 170k, 
37k, and 170k high-quality {\spz}, {\gprz}, and many-band {\phz}'s, respectively.

As described in \citet{tanaka+18}, to perform accurate cross-validation at {\hsc} \texttt{Wide} depths, 
each object was assigned an ``emulated'' \texttt{Wide}-depth flux error based 
on the error properties of similar objects 
observed in the {\hsc} \texttt{Wide} survey. Objects were also assigned an associated 
color-magnitude weight using a nearest-neighbor approach based on a representative subset of 
the weak lensing source sample to account for domain mismatches following the methodology described in
\S\ref{subsec:prior}. The original and re-weighted redshift distributions 
of the {\hsc} \texttt{S16A} training sample are shown in Figure \ref{fig:training}.

\section{How Representative are Existing Spectroscopic Redshift Samples?}
\label{sec:som}

As discussed in \S\ref{subsec:train}, spectroscopic ``completeness'' within our 
training set varies strongly across magnitude and color. In other words, in a given
color-magnitude ``bin'' the fraction of objects that come from more reliable sources such as \spz's versus 
more unreliable sources such as many-band \phz's can change rapidly.

This behavior is concerning for several reasons. First, {\spz}'s generally have much 
smaller (often negligible) errors in redshift measurements compared to {\phz}'s, so our underlying
knowledge of the redshift distribution at fixed magnitude and/or color degrades as the number
and/or fraction of {\spz}'s decreases. Second, it is a well known issue that {\phz} PDFs 
can be mis-calibrated (see, e.g., \citealt{tanaka+18}). Third, there may be systematic 
biases of the redshift distribution of {\spz}'s relative to {\phz}'s in a given color-magnitude bin
arising from selection effects that arise during the process of data collection. These involve
choices that often generate the mismatch in the first place, from prioritizing 
spectroscopic targets (which often impose magnitude and color biases) to how non-detections are treated
(which correlates strongly with redshift).

In particular, many studies assume that these pathological behaviors
can be ``calibrated out'' by matching objects explicitly in terms of color (not magnitude)
to obtain a representative {\spz} sample \citep[see, e.g.,][]{lima+08}. 
Surveys such as the Complete Calibration
of the Color-Redshift Relation (C3R2) Survey \citep{masters+17} 
have expanded upon this strategy, explicitly sorting possible targets into bins in color
space and then pursuing them assuming that {\spz}'s obtained at fixed color are representative of the
entire photometric population in that given color bin.

\begin{figure*}
\begin{center}
\includegraphics[width=\textwidth]{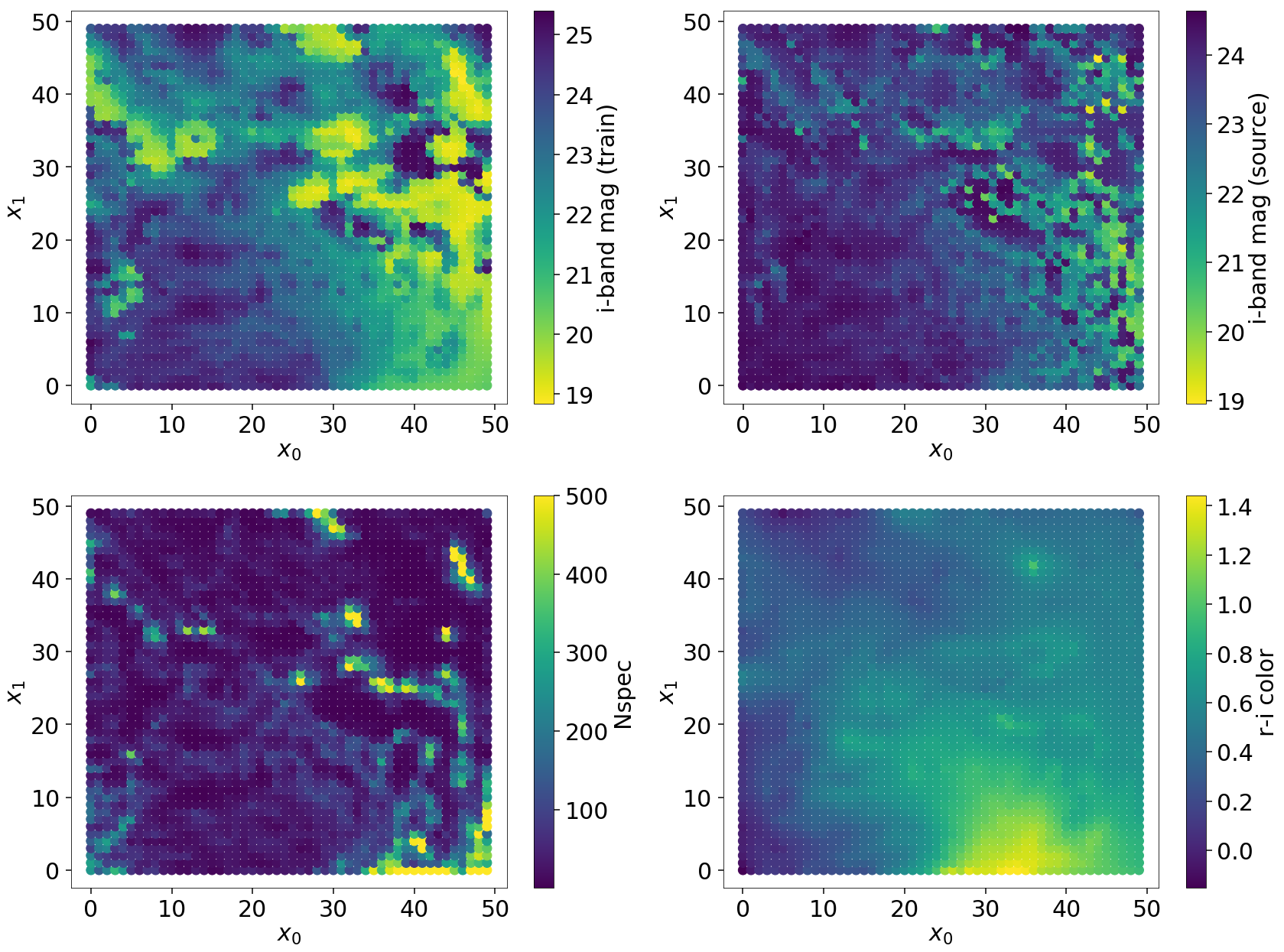}
\end{center}
\caption{The mean $i$-band magnitude of the training set (top left) 
and a randomly-selected subset of the 
weak lensing source catalog (top right),
the number of {\spz}'s (capped at 500; bottom left), 
and $r-i$ color (bottom right) of the $50 \times 50$ 
Self-Organizing Map (SOM) trained on the {\hsc} \texttt{S16A} weak lensing photometric sample. 
The {\spz}'s in our training sample occupy almost entirely mutually-exclusive 
regions of magnitude space
from the weak lensing photometric sample and have preferentially brighter magnitudes. 
As with Figure \ref{fig:training}, we see that large portions 
of color/magnitude space do not have sufficient
coverage within our dataset, necessitating the use of many-band {\phz}'s 
from surveys such as COSMOS to
``bridge the gap'' when computing {\phz}'s for {\hsc}.
}\label{fig:som}
\end{figure*}

While this strategy is efficient, it assumes that the intrinsic redshift distribution $P(z|\mathbf{c})$
at fixed color $\mathbf{c}$ is representative over all relevant magnitudes $m$. This is a strong assumption
given that the population of galaxies evolves as a function of redshift and that we expect
brighter objects of fixed color to be (on average) at lower redshifts, all else being fixed. 
We begin by investigating to what degree $P(m_{\rm spec}|\mathbf{c})$ differs 
from $P(m|\mathbf{c})$ and whether or not these differences are of importance to current lensing surveys.

More formally, this assumption implies that
\begin{align}
P(z|\mathbf{c}) &= \int P(z|\mathbf{c},m) P(m|\mathbf{c}) dm \\
&\approx \frac{1}{N_{\rm spec}} \sum_i K(z|z_{i, {\rm spec}}) \label{eq:spz_pz}
\end{align}
where $K(z|z_{i, {\rm spec}})$ is a kernel density estimate (i.e. smoothing scale) 
for each {\spz} where $z_{i, {\rm spec}}$ is a spectroscopic redshift drawn 
from a particular {\spz} distribution $P(m_{\rm spec}|\mathbf{c}) \neq P(m|\mathbf{c})$ 
based on how the data were collected. Note that, in general, \eqref{eq:spz_pz} 
is only guaranteed to be approximately valid if all the {\spz}'s comprise a 
representative sample from the underlying magnitude distribution
at fixed color $P(m|\mathbf{c})$.

To investigate these potential issues in our training sample, we will use 
manifold learning to sort our training galaxies into regions of color space 
to investigate possible trends in $P(z|\mathbf{c},m)$. In \S\ref{subsec:manifold}, we describe the
particular algorithm and procedure used to construct the manifold. In \S\ref{subsec:magcolor}, 
we examine several examples of $P(z|\mathbf{c},m)$ and find that there can be significant evolution
as a function of magnitude in particular regions of color. Our results imply that current {\spz} follow-up
programs should be cognizant of these effects in order to avoid biasing
{\phz} predictions at fixed color. Since the redshift success rate for a given spectrograph may depend on the redshift, at a given magnitude and color bin even surveys that are relatively homogeneously selected may be
subject to these subtle biases.

\subsection{Manifold Learning and Self Organized Maps}
\label{subsec:manifold}

For this study, we use a Self-Organizing Map \citep[SOM;][]{kohonen82,kohonen01}
to both cluster our data in color space and learn a lower-dimensional 2-D projection that can be used for
visualization purposes. We summarize the main features of our specific implementation below,
and direct the reader to \citet{carrascokindbrunner14}, \citet{masters+15,masters+17}, and
\citet{speagleeisenstein17a} for more details concerning their applications to {\phz}'s.

The SOM is an unsupervised machine learning algorithm that projects high-dimensional data 
onto a lower-dimensional space using competitive training of a (large) set of ``nodes'' in a
way that attempts to preserve general topological features and correlations present
in the higher-dimensional data. It consists of a fixed number of nodes
$N_{\rm nodes} = \prod_{i=1}^{D} N_{\rm nodes}^i$, where the product over $i$ is taken over all
dimensions $D$ of the SOM, arranged on an arbitrary $D$-dimensional grid with $N_{\rm nodes}^i$ nodes
in each dimension. Each node in the grid is assigned a position $\mathbf{x}$ on the SOM and is represented
by a particular model $\mathbf{F}(\mathbf{x})$, where $\mathbf{F}$ is the set of observed features.
In this paper, these are the set of $grizy$ photometric flux densities 
comprising a particular galaxy Spectral Energy Distribution (SED) in the HSC filters.

Once the dimensions/shape of the SOM are chosen, training then proceeds as follows:
\begin{enumerate}
\item Initialize the node models (most often randomly) and set the current iteration t = 0.
\item Draw (with replacement) a random object $\hat{\mathbf{F}}$ and its 
associated errors $\boldsymbol{\sigma}$ from the input dataset.
\item Compute 
\begin{equation}
\chi^2(\mathbf{x}) = \sum_b \sigma_b^{-2} (\hat{F}_b - s(\mathbf{x}) F_b(\mathbf{x}))^2
\end{equation}
across all nodes in the SOM over the available bands indexed by $b$, 
where the scale factor
\begin{equation}
s = \frac{\sum_b \sigma_b^{-2} \hat{F}_b F_b(\mathbf{x})}{\sum_b \sigma_b^{-2} F_b^2(\mathbf{x})}
\end{equation}
renormalizes the model SED so that we are fitting in terms of flux
density \textit{ratios} (i.e. colors) rather than flux densities (i.e. magnitudes) directly.
\item Select the best-matching node
\begin{equation}
\mathbf{x}_{\rm best} = \underset{\mathbf{x}}{\mathrm{argmin}} 
\left\lbrace \chi^2(\mathbf{x}) \right\rbrace
\end{equation}
based on the minimum $\chi^2(\mathbf{x})$ value.
\item Update the node models $\mathbf{F}(\mathbf{x}) \rightarrow \mathbf{F}^\prime(\mathbf{x})$ 
based on a learning rate $\mathcal{A}(t)$ and 
neighborhood function $\mathcal{H}(\mathbf{x}, \mathbf{x}_{\rm best}|t)$
such that 
\begin{equation}
\mathbf{F}^\prime(\mathbf{x}) = s(\mathbf{x}) \mathbf{F}(\mathbf{x}) 
+ \mathcal{A}(t) \mathcal{H}(\mathbf{x}, \mathbf{x}_{\rm best}|t) 
(\hat{\mathbf{F}}_i - s(\mathbf{x}) \mathbf{F}_i(\mathbf{x}))
\end{equation}
\item If $t \leq N_{\rm iter}$, increment $t$ and repeat starting from (ii).
\end{enumerate}

After training, objects are typically ``mapped'' onto the SOM by repeating steps (iii) and (iv)
for every object in the input dataset, assigning each object to its best-matching node.
We use a modified version of this approach where each object $\hat{\mathbf{F}}_j$ is
instead assigned to a set of nodes along with its corresponding weight
$w_j(\mathbf{x}) \propto e^{-\chi^2(\mathbf{x})/2}$ for all nodes with 
$w_j(\mathbf{x}) > f_{\min} \max(w_j(\mathbf{x}))$. 
We take $f_{\min} = 10^{-3}$, which approximately corresponds to thresholding
galaxies that are $\sim 2.5\sigma$ away from the best-fit.
This ``probabilistic mapping''
allows us to better capture the
uncertainty in an individual object's position on the SOM based 
on its photometric errors, resulting in smoother maps that are less 
sensitive to sampling noise and photometric errors relative to, e.g., \citet{masters+17}. 
A detailed discussion of these differences is beyond the scope of this paper
and will be explored in future work.

\begin{figure}
\begin{center}
\includegraphics[width=0.45\textwidth]{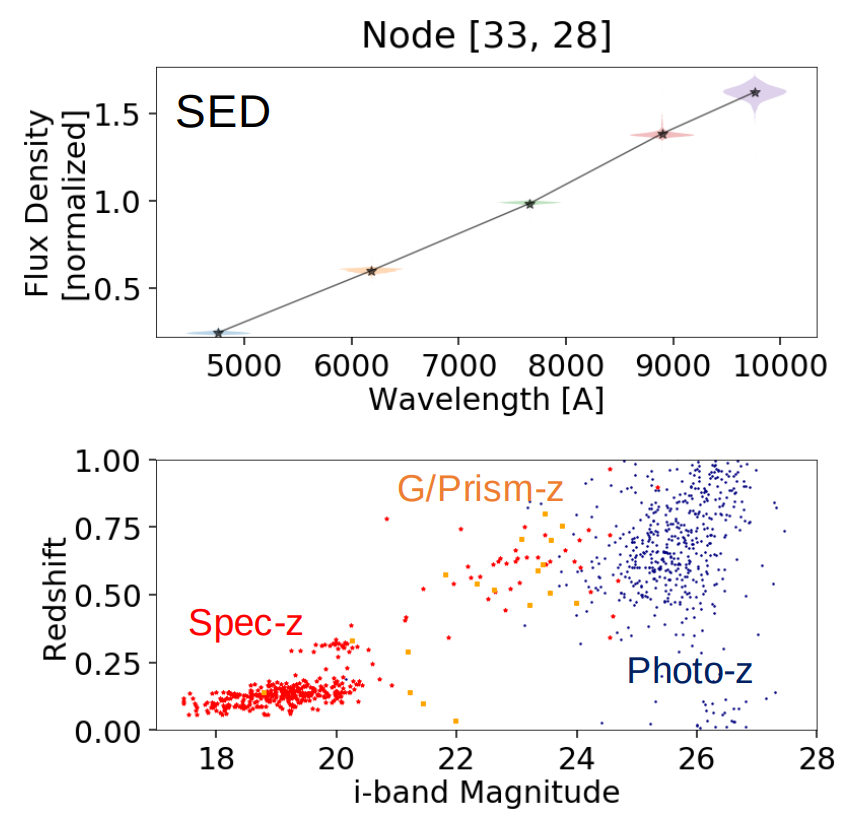}
\end{center}
\caption{\textit{Top:} The SED associated with a particular node in the SOM
located at position $\mathbf{x} = [33, 28]$.
The rescaled flux densities of objects associated with the node are shown as ``violin plots'' 
around the node SED, weighted according to each object's relative likelihood of being
associated with the node.
\textit{Bottom:} A Monte Carlo realization of the
corresponding redshifts of \spz's (red), \gprz's (orange), 
and many-band \phz's (blue)
and plotted based on their relative likelihood of being associated with the node.
We see a clear trend towards higher redshift as a function of magnitude. While some of the evolution
at fainter magnitudes is due to intra-bin scatter from photometric uncertainty, this effect
is minimal at brighter magnitudes where the trend is clearest. 
}\label{fig:som2}
\end{figure}

\begin{figure}
\begin{center}
\includegraphics[width=0.45\textwidth]{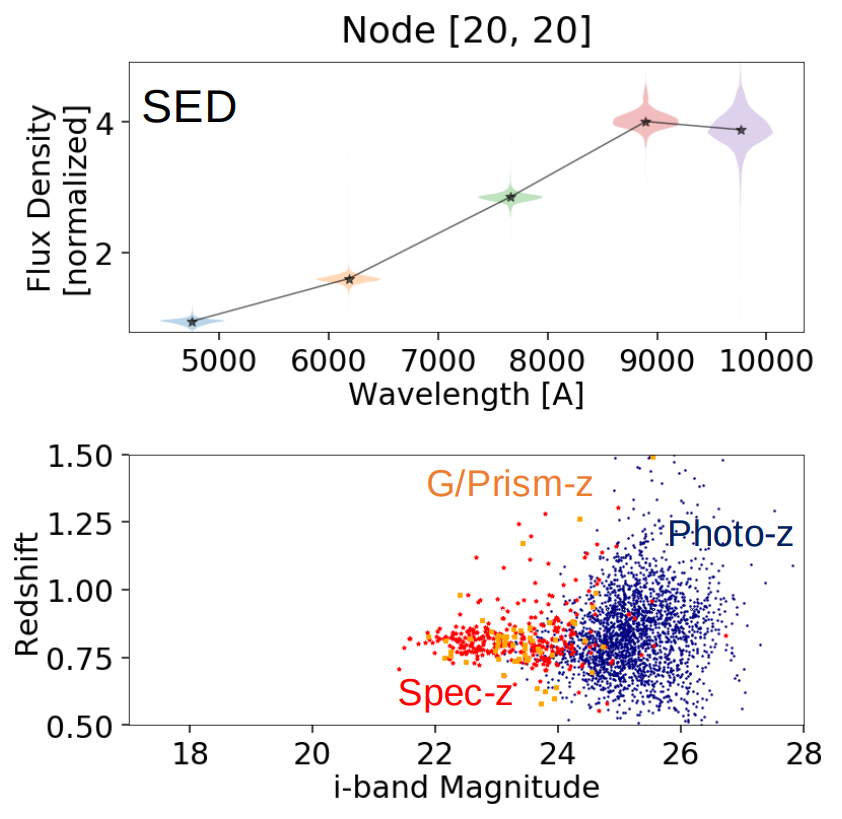}
\end{center}
\caption{\textit{Top:} As Figure \ref{fig:som2}, but for a particular node located at position 
$\mathbf{x} = [20, 20]$ in the SOM.
\textit{Bottom:} Monte Carlo realizations from the redshift PDFs of sources associated with the
node following Figure \ref{fig:som2}. In contrast to \ref{fig:som2}, for these
particular colors at most magnitudes the
overall behavior is consistent with having a non-evolving intrinsic redshift distribution.
}\label{fig:som3}
\end{figure}

We choose our SOM to be 2-D with a $50 \times 50$ grid of nodes, and train it
on the weak lensing source catalog photometry following the steps above
for $N_{\rm iter}=10^5$ iterations, which we find is enough to ensure the median
$\chi^2(\mathbf{x}_{\rm best})$ value for objects across our map 
is approximately the number of colors (four) available.
We choose our learning rate to be the weighted harmonic mean
\begin{equation}
\mathcal{A}(t) = \frac{1}{(1-t/N_{\rm iter})A_0^{-1} + (t/N_{\rm iter}) A_1^{-1}}
\end{equation}
for $A_0 = 0.5$ and $A_1=0.1$ and the neighborhood function to be a Gaussian kernel
\begin{equation}
\mathcal{H}(\mathbf{x}, \mathbf{x}_{\rm best}|t) 
= \exp\left( - 0.5 \, \sigma^2(t)\sum_i (x_i - x_{{\rm best},i})^2 \right)
\end{equation}
with a standard deviation that goes as the weighted harmonic mean
\begin{equation}
\sigma(t) = \frac{1}{(1-t/N_{\rm iter})\sigma_0^{-1} + (t/N_{\rm iter}) \sigma_1^{-1}}
\end{equation}
with $\sigma_0=35$ and $\sigma_1=1$. Our final SOM is shown in Figure \ref{fig:som}.

\subsection{Redshift Evolution at Fixed Color}
\label{subsec:magcolor}

Using our SOM, we can now investigate the questions outlined in the beginning of this section. In
particular, we want to examine whether the \textit{intrinsic} redshift distribution at fixed color
$P(z|\mathbf{c})$ is insensitive to magnitude within our training data. 

Although not common across our SOM, we do find some regions where there is evolution
in $P(z|\mathbf{c},m)$ at brighter magnitudes within {\spz}-dominated samples. One such example is shown
in Figure \ref{fig:som2}. 
For contrast, a more ``typical'' node shown for contrast in Figure \ref{fig:som3}. 
This confirms our basic intuition, formalized in Bayesian {\phz} approaches such as BPZ \citep{benitez00},
that the complicated evolution of galaxy SEDs and number densities as a function of time can
lead to $P(z|\mathbf{c},m)$ evolution as a 
function of magnitude if the underlying SED cannot be uniquely
constrained. While this is likely possible in future multiwavelength datasets 
with full optical to near-infrared coverage \citep[see, e.g.,][]{hemmati+18}, 
this likely remains a problem for current/planned weak lensing-oriented surveys such as 
{\hsc}, DES, KiDS, and LSST.

Note that we do expect that noisy observations will naturally lead to a broadening of the 
redshift distribution at fainter magnitudes due to intra-bin scatter (i.e. an object's PDF
gets ``smeared'' across multiple nodes on the SOM), and possibly to one whose mean
distribution evolves strongly with magnitudes, mimicking a shift in the intrinsic $P(z|\mathbf{c},m)$
distribution as a function of $m$.\footnote{The shift in the mean can be due to asymmetric 
scattering caused by secondary redshift solutions and changing number densities in color space.} 
This effect, however, should not impact
the redshift distribution at brighter magnitudes (where measurement errors
are small), which is where most of our {\spz}'s lie and where the trend seen in Figure \ref{fig:som2} is the most apparent.

To quantify the extent to which possible
redshift evolution can impact our redshift results, we focus on evolution
in {\spz} observations at $i$-band magnitudes brighter than $m=22.5$ to mitigate
redshift errors based on photometric measurement errors and incorrect redshift solutions.
We compute the median redshift in bins of 0.5 mag, and fit linear trends for
all SOM nodes where we could compute medians for
$\geq 3$ bins using $\geq 10$ {\spz} observations in each bin. 
We find that of the 709 nodes which fit this criteria ($\approx 30\%$ of the SOM), 
around 40\% (287) display significant redshift evolution with $dz/dm > 0.05$.
This trend is robust to different choices in $dz/dm$ threshold and the required
number of bins used in the fit, and is substantially higher than the few percent
expected due to random variation. While this test is limited in scope,
it highlights that the trend shown in Figure \ref{fig:som2} is not an isolated case
and needs to be taken seriously.

These results indicate that it can be dangerous to use re-weighted {\spz}
samples based on only a few broadband colors and expect to get the correct
$P(z|\mathbf{c})$ distribution, a danger which is indeed recognized by other weak lensing analyses
\citep[e.g.,][]{troxel+17,kohlinger+17,hikage+18}. 
\textit{This implies that {\spz} samples may need to be 
representative in magnitude as well as color.}\footnote{This does not address the secondary issue of 
redshift-dependent failure rates at a fixed magnitude and color, which will also bias the resulting sample.}

\section{Photometric Redshift Framework}
\label{sec:photoz}

Based on the results in \S\ref{sec:som}, we aim to develop a framework that allows
us to explicitly incorporate magnitude-dependence into our {\phz} predictions to probe $P(z|\mathbf{c},m)$
and alleviate possible mismatches at fixed color between {\spz}'s and many-band {\phz}'s
present within our training set. At fainter magnitudes, however, almost all objects
that make up our estimates for $P(z|\mathbf{c},m)$ come from many-band {\phz}'s.
As a result, we also want to track how \textit{individual} objects in the training set propagate forward
to our eventual {\phz} predictions to investigate how much our many-band {\phz}'s are contributing to 
redshift predictions in different regions of color-magnitude space.

We adopt a Bayesian-oriented nearest-neighbors (NN)-based approach 
that attempts to properly account for measurement errors within both training and testing sets 
when making {\phz} predictions based \textit{explicitly} on observed flux densities (magnitudes). 
In \S\ref{subsec:bayes}, we discuss the Bayesian underpinning of our approach. 
We describe our likelihood in \S\ref{subsec:like} and our
NN-based approximations to the likelihood/posterior in \S\ref{subsec:knn}. 
We discuss our priors in \S\ref{subsec:prior}.

All photometric redshifts (and SOMs) in this study were computed using an early development version
(\texttt{v0.1.5}) of the Python {\phz} package \texttt{frankenz}\footnote{Available 
online at \url{github.com/joshspeagle/frankenz}.} (Speagle et al. in prep.).

\subsection{Bayesian Inference}
\label{subsec:bayes}

Deriving photometric redshifts ultimately relies on modeling the \textit{continuous} mapping 
between a set of observables $\mathbf{F}$ (e.g. flux densities) within some number of 
bands and redshift $z$. The central idea of our approach
is that in the ``big data'' limit
this comparison can instead be approximated as a \textit{discrete} 
comparisons between individual objects. 
The redshift for a target galaxy indexed by $g$ out of $N$ galaxies given a set of $M$ training
galaxies indexed by $h$ can then be written as
\begin{equation}\label{eq:phz}
P(z|g) = \sum_{h=1}^{M} P(z|h) P(h|g) = \sum_{h=1}^{M} P(z|h) \frac{P(g|h) P(h)}{P(g)}
\end{equation}
where $P(z|h)$ is the redshift PDF for galaxy $h$, $P(h|g)$ is the 
posterior, $P(g|h)$ is the likelihood, $P(h)$ is the prior for $h$, and 
$P(g) = \sum_{h=1}^{M} P(g|h)P(h)$ is the evidence (marginal likelihood). 
In other words, we are trying to find 
the probability that our observed galaxy $g$ is actually a realization of our training galaxy $h$ 
(i.e. whether $g$ and $h$ are a photometric ``match''). We then assign it the corresponding 
redshift kernel $P(z|h)$ for galaxy $h$ with a weight proportional to the posterior probability $P(h|g)$. 
$P(z|g)$ then becomes a posterior-weighted mixture of our $P(z|h)$'s.

\begin{figure*}
\begin{center}
\includegraphics[width=\textwidth]{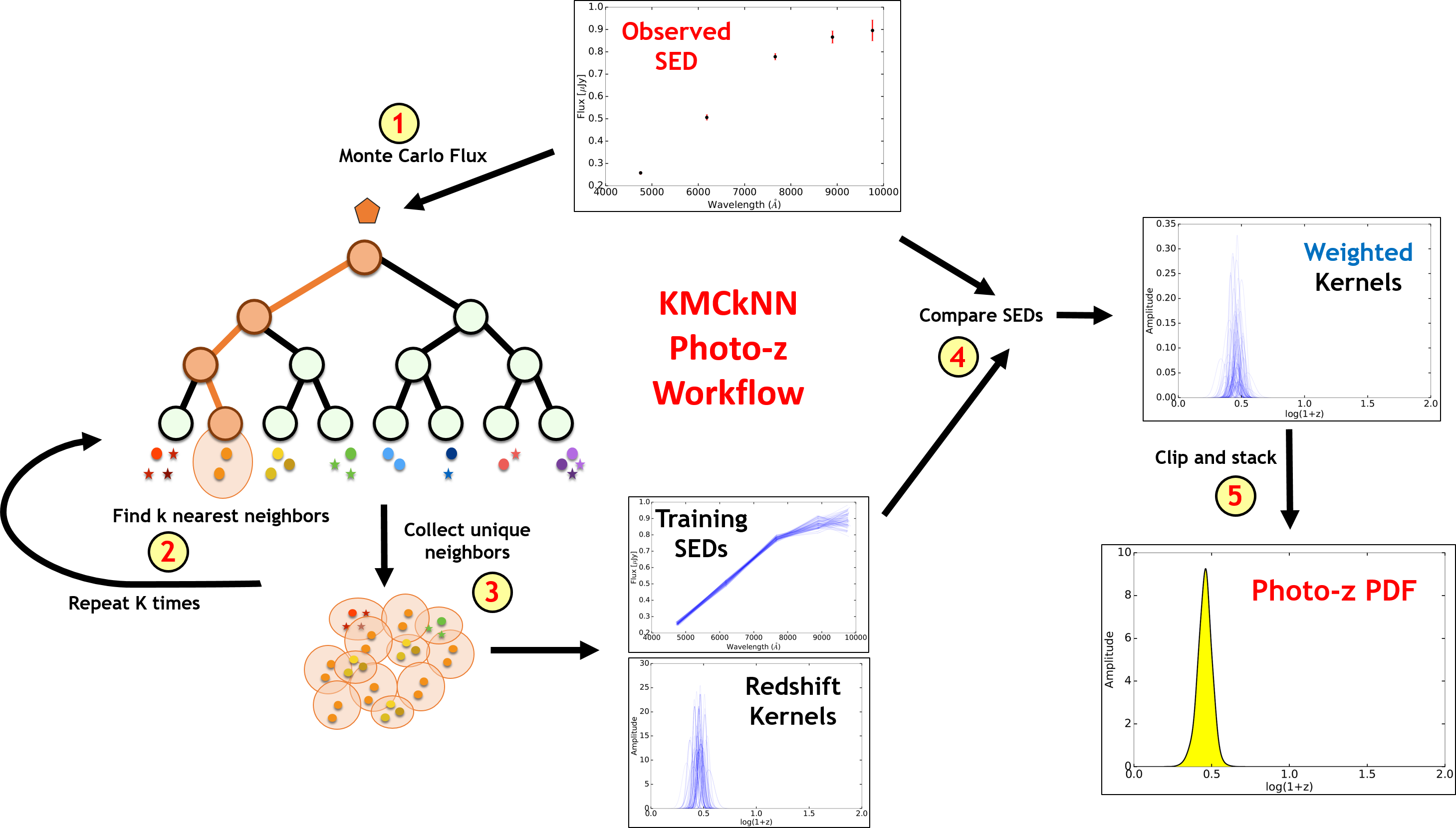}
\end{center}
\caption{A schematic outline of the major steps in our KMCkNN based algorithm
described in \S\ref{subsec:knn}. First, a Monte Carlo realization of the target object 
indexed by $g$ is matched with a set of $k$ nearest neighbors based on a 
Monte Carlo realization of all training objects indexed by $h$. After
repeating this procedure $K$ times, all unique neighbors in the training set 
are identified and the corresponding photometric posteriors $P(h|g)$ are computed.
These are then used to derive a {\phz} PDF based on a posterior-weighted mixture of the corresponding
redshift kernels $P(z|h)$.
}\label{fig:kmcknn}
\end{figure*}

\subsection{Photometric Likelihood}
\label{subsec:like}

Assuming the errors on the measured fluxes $\mathbf{\hat{F}}_g$ and $\mathbf{\hat{F}}_h$ are 
independent and Normal (i.e. Gaussian) and ignoring the impact of selection effects (see, e.g.,
\citealt{leistedt+16}), the log-likelihood for $P(g|h)$ from our set of $B$ bands indexed by $b$
can be naively written as 
\begin{align}
-2 \ln P(g|h) &= \sum_{b=1}^{B} \frac{\left( \hat{F}_{g,b} - \hat{F}_{h,b} \right)^2}
{\hat{\sigma}^2_{g,b} + \hat{\sigma}^2_{h,b}} + \ln(\hat{\sigma}^2_{g,b} + \hat{\sigma}^2_{h,b}) \\
&+ B\ln 2\pi \nonumber
\end{align}
This represents the standard $\chi^2$ statistic often used in template-fitting methods 
(see \S \ref{subsec:manifold}) but with error contributions from both the target ($g$)
and training ($h$) objects and including the relevant normalization term.

Unlike most template-fitting codes and contrary to the approach taken in \S\ref{sec:som}, 
we deliberately chose \textit{not} to include a free scaling parameter $s$ 
to try and account for normalization offsets between $g$ and $h$ (i.e. fitting in magnitudes instead of 
colors). There are two reasons for this. The first is that
the conditional prior $P(s|h)$ we would want to impose over 
$s$ when computing $P(g|h) = \int P(g|h,s)P(s|h) ds$ is unclear. For instance, a fixed 
prior such as the uniform $P(s|h) = P(s) = 1$ prior used by most template-fitting codes 
is equivalent to assuming a fixed (and unphysical) luminosity function, 
which can create biases in inference. 
Trying to specify a color-dependent luminosity function $P(s|h)$ directly, however, 
is extremely challenging because the integral over $s$ often cannot be evaluated analytically.
Fitting flux densities $\mathbf{F}$ directly avoids these complications.

More importantly for our purposes, however, is that our results from \S\ref{sec:som} show that there
can be strong evolution in the underlying redshift distribution $P(z|\mathbf{c}, m)$ at fixed color
$\mathbf{c}$ as a function of magnitude $m$. 
To mitigate this effect without attempting to deal with complicated priors,
for simplicity we opt to keep the likelihood ``close to the data'' and fit directly in $\mathbf{F}$.

\subsection{Nearest-Neighbors Approximation}
\label{subsec:knn}

To avoid running over all $M$ objects in the training set, we use a modified nearest-neighbors approach
to preferentially select objects that have similar flux densities with respect to their errors.
As the relative errors $\hat{\boldsymbol{\sigma}}^2_g + \hat{\boldsymbol{\sigma}}^2_h$ 
of any two training/target objects $g$ and $h$ will differ, the relevant distance metric 
will be different for every pairwise training-target object combination. 
As nearest neighbor searches are typically done with respect to a fixed distance metric (often
the Euclidean distance), this pairwise distance dependence poses a problem.

We deal with this by using Monte Carlo methods to search for neighbors across 
multiple realizations of the observed flux densities. We first
generate a Monte Carlo realization $\tilde{\mathbf{F}}_g$ and $\tilde{\mathbf{F}}_h$ 
of the photometry for all objects in our target set and training set, respectively.
We then determine the $k$ nearest
neighbors based on the Euclidean squared distance between our set 
of Monte Carlo-ed flux densities
to a given observed galaxy $g$ using a $k$-$d$ tree \citep{bentley75}, defining
a set of $k$ indices $\tilde{\mathbf{h}}(g)$. After repeating this process $K$ times, 
we define an object's set of 
``photometric neighbors'' as the \textit{union} of the $k$ nearest 
neighbors from each of the $K$ Monte Carlo realizations:
\begin{equation}
\tilde{\mathbf{H}}(g) = \tilde{\mathbf{h}}_1(g) \cup \cdots \cup \tilde{\mathbf{h}}_K(g) 
\end{equation}

Using our $K$ Monte Carlo $k$ nearest neighbors (KMCkNN) approximation, we
only compute photometric likelihoods to a small fraction of the data preferentially
selected to contain the highest likelihoods. This procedure is most robust when the set of neighbors
is roughly complete out to 3-5 standard deviations (relative to all possible 
pairwise galaxy combinations). An object can have at most $k \times K$ possible neighbors,
with the exact number a strong function of the signal-to-noise of the target object $g$ 
and the density of training objects $h$ in its local region of color-magnitude space. 

The sparse ($kK \ll N_{\mathbf{h}}$) nature of this KMCkNN approximation enables
us to keep track of the \textit{individual} log-likelihoods $\ln P(g|h)$ 
and their corresponding indices $\tilde{\mathbf{H}}(g)$
across all target objects. Because these have been computed exclusively using observables, 
we can subsequently use them to construct \textit{any} associated ancillary quantities
``after the fact''. The most relevant example is the {\phz} PDFs following equation \eqref{eq:phz}, 
but this may include a whole range of other useful quantities such as those
detailed in \S\ref{subsec:pznew}. 
A schematic diagram of our KMCkNN approach is shown in Figure \ref{fig:kmcknn}.

One significant drawback of using a 
nearest-neighbors approach is that it is difficult to accommodate 
missing data. However, since the weak lensing source catalog used in this 
work is only defined in regions with
full depth and full color coverage \citep{mandelbaum+18}, this restriction
does not impact our results.

\begin{figure*}
\begin{center}
\includegraphics[width=\textwidth]{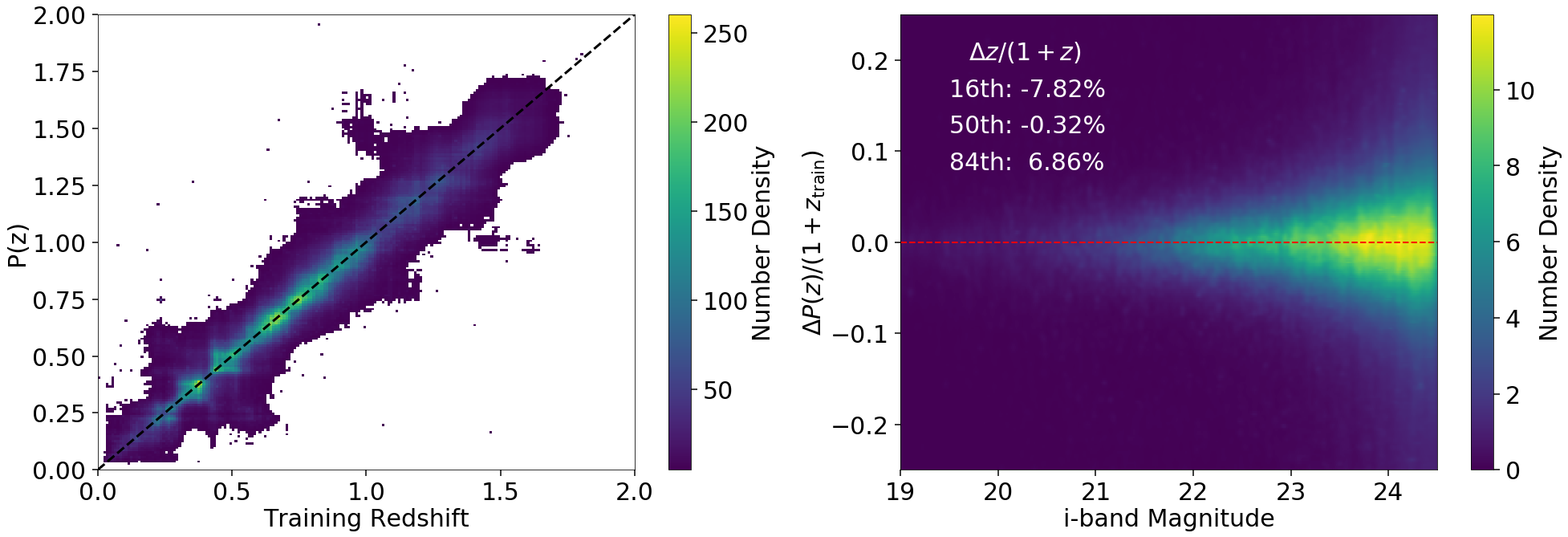}
\end{center}
\caption{\textit{Left:} The 2-D stacked distribution of our cross-validation
{\phz} PDFs as a function of the
training redshift for the re-weighted training samples based on the
{\hsc} \texttt{Wide}-depth emulated errors (see \S\ref{subsec:train}).
\textit{Right:} The reweighted redshift-normalized dispersion $\Delta z / (1+z)$ of our
{\phz} PDFs relative to the training redshifts as a function of $i$-band magnitude. 
The 16th, 50th, and 84th percentiles of the marginalized redshift-normalized dispersion 
for the sample are shown in the upper portion of the figure. 
Our typical 68\%-level uncertainty is $\approx 7.35\%$ with a median bias of $-0.3\%$, 
illustrating that our {\phz} PDFs are robust and unbiased with respect to our training data.
}\label{fig:pz1}
\end{figure*}

\subsection{Photometric Priors}
\label{subsec:prior}

We incorporate the KMCkNN approximation from \S\ref{subsec:knn} 
into our photometric prior $P(h)$ by defining
a new ``sparse'' prior
\begin{equation}
\tilde{P}_g(h) =
\begin{cases}
P(h) \quad &\text{for} \: h \in \tilde{\mathbf{H}}(g) \\
0 \quad &\text{otherwise}
\end{cases}
\end{equation}
Our {\phz} PDFs can then be written as
\begin{equation}\label{eq:phz_sparse}
P(z|g) \approx \sum_{h \in \tilde{\mathbf{H}}(g)} P(z|h) \frac{P(g|h) P(h)}{P(g)}
\end{equation}

Typically, the prior $P(h)$ is defined to adjust for ``domain mismatch''
between the training and target
datasets. Since {\spz} training sets are significantly biased in both 
color and magnitude relative to most
target photometric galaxy populations, this ``reweighting'' via $P(h)$ traditionally 
substantially improves {\phz} accuracy relative to cases where $P(h)$ is assumed
to be uniform \citep{lima+08}.

We compute a photometric ``prior'' using the magnitude-based, KMCkNN approach 
described in \S\ref{subsec:like} and \S\ref{subsec:knn} by computing the 
approximate Bayesian evidence under a uniform prior
\begin{equation}
P(h) \approx \sum_{g \in \tilde{\mathbf{G}}(h)} P(h|g)
\end{equation}
where our roles for $g$ and $h$ have switched: we now treat our set of
training galaxies indexed by $h$ as ``target'' objects and a subsample of
$N_{\rm ref}$ reference objects indexed by $g$ as ``training'' objects. 
We found the impact of the prior on our redshift predictions
in internal testing was mostly unchanged for values of
$N_{\rm ref} \gtrsim N_{\rm train}$, and so opt to use $N_{\rm ref} = \sn{5}{5}$
here.

To put this procedure another way,
we determine $P(h)$ by stacking the likelihoods of neighbors in the target data 
around individual training objects. This procedure, while not entirely proper
(we are using a subset of galaxies in our sample to determine the prior),
is sufficient for our purposes and represents the first step to a proper
hierarchical model \citep{leistedt+16}.

We find our prior-weighted training data are able to
reproduce the $B$-dimensional distribution of our target data quite well as long 
as $K$ and $k$ are sufficiently large.
See \citet{tanaka+18} for additional details.

\subsection{New Quality Indicators}
\label{subsec:pznew}

Unlike other machine learning-oriented approaches, we are able to compute (approximate)
posterior quantities to \textit{every} training-target object pair. 
This enables us to utilize a variety 
of Bayesian-oriented indicators to determine the quality of our fits. 
We will discuss two new quality indicators here: metrics related to basic goodness-of-fit tests 
(\S \ref{subsubsec:quality}) and those describing the ``information content'' used in our predictions
(\S \ref{subsubsec:fspec}).

\subsubsection{Goodness-of-Fit}
\label{subsubsec:quality}

By computing posterior quantities to every pair of training-target objects, we can
exploit goodness-of-fit tests used in a broad set of Bayesian model fitting applications. 
We will examine the two most basic indicators here: the maximum a posteriori (MAP) result 
$\rom{P}{max}(g) \equiv \max \lbrace \dots, P(h|g), \dots \rbrace$ and the evidence 
$P(g) \equiv \sum_{h \in \mathbf{h}} P(g|h)P(h)$. 

The MAP quantifies how good our best-fit result is, enabling us to determine if a given set of observables is
represented in our training data. This is extremely useful when trying to remove objects
with unreliable predictions that lie outside the parameter space spanned by our training data.

The evidence quantifies how well an object is represented across the entirety of our training data. 
This is useful when trying to identify objects which do not have meaningful ``coverage'' since 
objects that are only similar to a handful of training
examples might have unreliable predictions.

In general, we find that the MAP and the evidence are highly correlated among our data: 
objects that are well-fit by at least one training example are very likely to be well-fit by others, 
and vice versa. Based on internal testing, \textit{we
find that instituting a cut explicitly on the best-fit result based on the fitted
$\chi^2$ values removes the majority of poorly-fit objects from our sample}.
Our final cut is based on the 95th quantile $P(\chi^2_5 \leq X) = 0.95$, 
where $\chi^2_5$ is the $\chi^2$ distribution
with 5 degrees of freedom, which is conceptually roughly equivalent to 2-sigma clipping.

\subsubsection{Information Content}
\label{subsubsec:fspec}

As mentioned in \S \ref{subsec:knn}, our sparse KMCkNN approximation allows 
us to keep track of individual
log-likelihoods computed between sets of training-target object pairs.
It is then straightforward to transform
these results into {\phz} PDFs via equation \eqref{eq:phz_sparse}.

More generally, however, keeping track of the relevant individual posterior predictions
\begin{equation*}
\tilde{P}_g(h|g) = \frac{P(g|h) \tilde{P}_g(h)}{\sum_h P(g|h) \tilde{P}_g(h)}
\end{equation*}
allows us to compute almost any
posterior-dependent result. This flexibility enables us to investigate auxiliary properties of interest.

In this work, we explore the impact of {\phz} systematics on weak lensing using our heterogeneous
{\hsc} training data. In particular, we are worried about the impact many-band {\phz}'s
might have on our results. 
In order to do this, \textit{we introduce two quantities to keep track of where the information 
content in a given {\phz} prediction $P(z|g)$ actually originates}:
\begin{itemize}
\item \texttt{Fphot}: the fraction of neighbors in the training set with many-band {\phz}.
\item \texttt{Pphot}: the \textit{posterior-weighted} fraction of neighbors in the training set with many-band {\phz}.
\end{itemize}

\textit{\texttt{Fphot} and \texttt{Pphot} will help us to determine what kind of redshift (e.g. photo-$z$, spec-$z$, ..) any given object has been trained on}. This will be an important ingredient for our lensing tests in Section \ref{sec:gglens}.



\section{Photometric Redshift Validation}
\label{sec:pzchecks}

In this section we outline the implementation (\S\ref{subsec:tune}), 
validation (\S\ref{subsec:cv}), and application (\S\ref{subsec:spec_complete}) 
of the {\phz} framework outlined in \S\ref{sec:photoz}.

\subsection{Tuning: Feature Selection and (Hyper-)parameter Choices}
\label{subsec:tune}

The {\hsc} catalog contains a variety of features that can be used for {\phz} predictions,
including a variety of photometry measurements and size information. 
In addition, the KMCkNN framework
described in \S\ref{subsec:knn} involves several hyper-parameters that can impact performance.

We conduct a variety of internal cross-validation and hold-out tests 
following \S\ref{subsec:cv} to determine the subset of features and 
hyper-parameters that give the best performance at a reasonable computational cost. 
Our results are summarized below:
\begin{itemize}
\item Our chosen flux density measurements were PSF-matched 1.1{\qq} aperture
photometry among objects with successful forced photometry in all five bands. Adding additional 
features such as size or using different combinations of other photometry products 
(e.g., \texttt{cmodel}) gave comparable or worse results.
\item We introduce a photometric smoothing kernel $\sigma_{g,b} = f_b \hat{F}_{g,b}$ 
for each object $g$ in each band $b$ to account 
for systematic uncertainties in measured photometric errors and to serve as a smoothing scale when
computing likelihoods. We find that $f_b = 0.02$ gives good {\phz} PDFs in aggregate and constitutes an
effective zero-point calibration uncertainty of $\approx 0.02 \,{\rm mag}$ 
(although see \citealt{tanaka+18}).
\item To ensure optimal runtime, 
we want to make $K$ and $k$ only as large as necessary to obtain 
good magnitude/color-space coverage for each object. 
The lower limit on $K$ is set by the number of Monte Carlo realizations 
needed to roughly marginalize over the measurement uncertainties 
when searching for neighbors, while the lower limit on $k$ is to 
ensure a reasonably large collection of neighbors. 
Based on internal testing, we find $K = 25$ Monte Carlo realizations 
and $k=10$ neighbors selected at each iteration works as a reasonable compromise.
The worst case performance ($N_{\rm nghbr} \sim 10$) generally only 
occurs for bright and rare objects, which usually also have poor likelihoods.
The typical number of unique neighbors is $N_{\rm nghbr} \sim 100-200$.
\item We take our redshift kernels to be Normal distributions 
$\Normal{z}{\mu=\hat{z}_h}{\sigma^2=(\Delta z)^2 + \hat{\sigma}_{h,z}^2}$ centered on the 
measured redshift $\hat{z}_h$ with a variance set by a combination of 
an intrinsic width $\Delta z = 0.01$, similar to the redshift spacing used when storing most {\phz} PDFs,
and the associated redshift measurement error $\hat{\sigma}_{h,z}$. 
This allows us to propagate uncertainties from the many-band {\phz}'s to our final predictions.
\end{itemize}

Following \citet{tanaka+18}, our redshift PDFs $P(z|g)$ are evaluated over a redshift grid ranging
from $0 \leq z \leq 6$ with $\Delta z = 0.01$ spacing. All redshift-based quantities described 
later in the text are derived from these discretized PDFs.

\begin{figure}
\begin{center}
\includegraphics[width=0.48\textwidth]{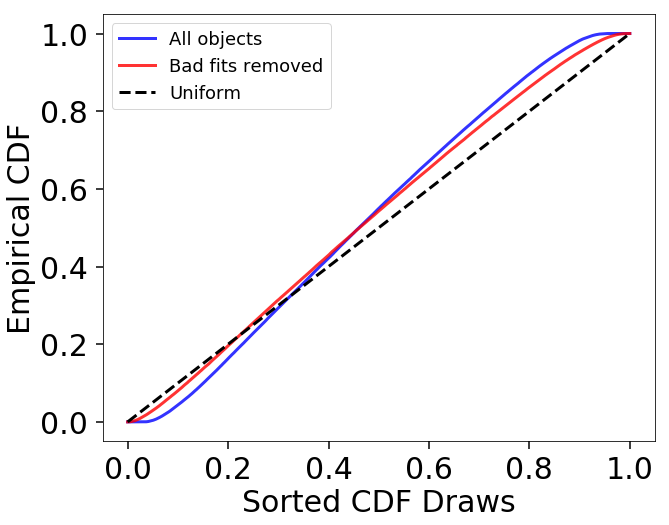}
\end{center}
\caption{The empirical cumulative distribution function (eCDF) of the ``true'' redshifts 
evaluated over the cross-validation {\phz} PDFs across the re-weighted training sample
before (blue) and after (red) removing ``bad'' fits ($\sim 5\%$ of the sample)
using the criteria outlined in \S\ref{subsubsec:quality}. 
The ideal behavior is shown as the dashed black line. When limiting to objects that are more
represented in our training sample (i.e. have better fits), we are able to remove most outliers with
poorly determined PDFs.
}\label{fig:pz3}
\end{figure}

\subsection{Calibration: Characterizing Behavior with Cross-Validation}
\label{subsec:cv}

As discussed in \S \ref{subsec:train}, to account for inhomogeneity and domain
mismatch in the training set all objects are assigned an
{\hsc} \texttt{Wide}-depth emulated error 
following the procedure described in \citet{tanaka+18} and an associated
color-magnitude weight following \S \ref{subsec:prior}. 
Unless stated otherwise, we utilize both quantities when computing any of the
performance estimates reported here.

\begin{figure*}
\begin{center}
\includegraphics[width=\textwidth]{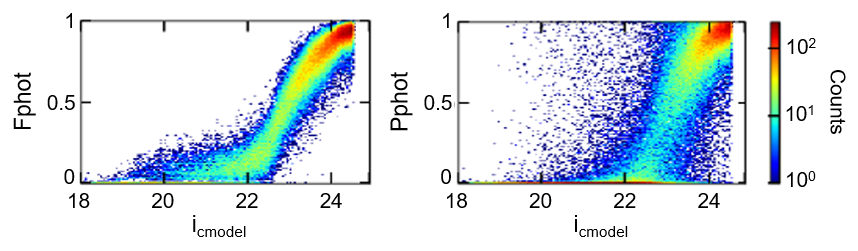}
\end{center}
\caption{The number density of \texttt{Fphot} (left) and \texttt{Pphot} (right) (see \S\ref{subsec:pznew})
as a function of $i$-band magnitude for a representative subsample of $\sim 300$k objects 
in our {\hsc} \texttt{S16A} weak lensing sample. As expected, \textit{the 
{\phz}'s of objects at $i > 23$ are trained almost entirely on the many-band 
{\phz}'s in our training sample}, while those at brighter magnitudes $i < 23$ tend to be trained on {\spz}'s and {\gprz}'s. This transition happens more smoothly in \texttt{Fphot} than \texttt{Pphot} because
the exponential nature of the likelihood tends to strongly favor a few photometric neighbors over
others (see \S\ref{subsec:like}). This behavior is most apparent at brighter magnitudes, where
even though $\sim 15\%$ of neighbors come from many-band {\phz}'s, they tend to contribute very little
to the overall {\phz} prediction.
}\label{fig:sel}
\end{figure*}

We first randomly divided our training data into 
validation/hold-out testing sets comprised of
$(1-f)$/$f$ fractions of the data for some hold-out fraction $f$.
We then use two strategies to select our hyper-parameters and evaluate our performance within
the validation set: $k=5$-fold cross-validation and
internal leave-one-out tests. For $k=5$-fold cross-validation, we randomly divided our validation
set into $k=5$ subsets. We then train on $k-1$ of these subsets to compute {\phz} predictions
to the remaining subset, cycling through each of the subsets until we had obtained predictions
to the entire validation sample. For leave-one-out tests, we instead train on the entire
validation set. However, when computing predictions to each object, we ``mask out''
its possible contribution within the selected group of neighboring objects used to
compute the {\phz} prediction. Both of these procedures, along with the final hold-out test
set, attempt to mitigate over-fitting and ensure realistic performance estimates.

We find that the results from \S \ref{subsec:tune} are mostly insensitive to the chosen 
hold-out fraction $f$ when $f \gtrsim 0.5$ 
(i.e. when our validation set consists of more than $\sim$\,150k objects). In addition, we
also find that the performance on the hold-out test set
is essentially identical to performance estimates within 
the validation set using both strategies when $f \gtrsim 0.8$,
confirming that our approaches avoid over-fitting and that the information content
appears to roughly saturate as our validation set exceeds $\sim$\,250k objects.
Based on these results, we find that it is reasonable to treat our features 
and hyper-parameters from \S \ref{subsec:tune} as
essentially fixed. Our reported performance is then estimated by applying
the more conservative $k=5$-fold cross-validation tests across the 
entire training sample (i.e. without the $(1-f)/f$ validation/testing split).

The 2-D stacked {\phz} PDFs versus the input true redshifts 
(smoothed by their intrinsic uncertainties)
along with the associated dispersion in $\Delta z / (1+z)$ as a function of
magnitude are shown in Figure \ref{fig:pz1}. 
We see that our performance over the weak lensing sample is
relatively robust, with an overall $\Delta z / (1+z) \approx -0.3\%$ bias
and with 68\% of the PDFs contained within $\Delta z / (1+z) = [-7.8\%, 6.9\%]$.

In addition to tests on the overall accuracy of our predictions, we also test the 
reliability of our individual
PDFs. We opt to use the empirical cumulative distribution function (eCDF),
which is constructed by evaluating the true redshift of each 
cross-validation object $i \in \mathbf{i}$ at the value of the predicted {\phz} CDF
\begin{equation}
\hat{u}_i \equiv \int_0^{z_i} P(z|i) dz
\end{equation}
and computing
\begin{equation}
\hat{U}(x) = \sum_i \mathbb{I}(\hat{u}_i \leq x)
\end{equation}
where $\mathbb{I}(\cdot)$ is the indicator function which returns $1$ if the condition is true
and $0$ if it is false. In the case where our PDFs are properly calibrated and
have the expected coverage provided by any associated confidence interval, each CDF draw 
$\hat{u}_i \sim \textrm{Unif}(0, 1)$ will be
uniformly distributed from 0 to 1 and $\hat{U}(x)$ will
approximately define a straight line from 0 to 1.

We show the eCDF results for our {\phz} PDFs in Figure \ref{fig:pz3}. 
These confirm that our PDFs are relatively
robust and internally well-calibrated.

\subsection{Application: Estimating Spectroscopic Incompleteness}
\label{subsec:spec_complete}

We now turn our attention to the motivating issue behind the development of 
the {\phz} framework outlined in \S\ref{sec:photoz} by investigating 
the distribution of \texttt{Fphot} and 
\texttt{Pphot} (see \S\ref{subsubsec:fspec}) within our {\hsc} {\phz}'s.

We show the distribution of \texttt{Fphot} and \texttt{Pphot} as a function of
magnitude in Figure~\ref{fig:sel}. The results are as expected: \textit{many-band {\phz}'s 
in our training sample make up an increasing large fraction
of neighbors and contribute an increasing amount to the {\phz} PDFs at fainter magnitudes}.
We will use \texttt{Fphot} and \texttt{Pphot} to investigate the robustness
of weak lensing measurements as a function of spectroscopic incompleteness (see \S\ref{sec:gglens}).

\section{Lensing Methodology}
\label{sec:ggtheory}

We now outline the methodology we will use to stack, compute, and compare our gg lensing signals 
based on the {\phz} PDFs illustrated in \S\ref{sec:pzchecks}.
We describe our basic computation of $\Delta \Sigma$ in \S\ref{subsec:ds} 
and our treatment of the bias/dilution factors in \S\ref{subsec:wlbias}.  We outline the approach
used to compare gg lensing signals between two samples in \S\ref{subsec:wl2samp}.

\subsection{Computing the Galaxy-Galaxy Lensing Signal}
\label{subsec:ds}

Our computation of the lensing observable, $\Delta \Sigma$, follows
the methodology of \cite{singh+17}. We use the code 
\texttt{dsigma}\footnote{Available online at \url{https://github.com/dr-guangtou/dsigma}.}, 
which was specifically written for computing gg lensing signals for {\hsc}.

We compute $\Delta \Sigma$ as a function of physical radius $R$ as

\begin{equation}
{\Delta\Sigma}_{\rm LR}(R) = 
f_{\rm bias}({\Delta\Sigma}_{\mathrm{L}}(R) - {\Delta\Sigma}_{\mathrm{R}}(R))
\label{eq:ds1}
\end{equation}

\noindent where ${\Delta\Sigma}_{\mathrm{L}}$ is the stacked signal around lens galaxies, 
${\Delta\Sigma}_{\mathrm{R}}$ is the stacked profile around a much larger number of random positions,
and $f_{\rm bias}$ is a correction factor (see \S\ref{subsec:wlbias}).
The ${\Delta\Sigma}$ profile for both lenses and randoms are computed as follows:

\begin{equation}
{\Delta\Sigma}_L(R) = \frac{1}{2 R(R) [1+K(R)]}
\frac{\Sigma_{\rm Ls}^{R} w_{\rm Ls} \gamma_{t}^{(\rm Ls)} 
\Sigma_{\rm crit}^{(\rm Ls)}}{\Sigma_{\rm Ls}^{R} w_{\rm Ls}}
\label{eq:ds2}
\end{equation}

\noindent where $\sum_{\rm Ls}^{R}$ indicates a sum over all lens-source pairs 
with separation $R$. When computing
${\Delta\Sigma}_R(R)$, we replace $\sum_{\rm Ls}^{R}$ with $\sum_{\rm Rs}^{R}$
since we instead sum over all random-source pairs. 

In Equation \ref{eq:ds2}, $\gamma_t$ is the tangential shear of a source galaxy:
\begin{equation}
\gamma_{\rm t} = - e_{1} \cos{2\phi} - e_{2} \sin{2\phi}
\label{eq:gamma}
\end{equation}
where $e_1$, $e_2$, and $\phi$ are the two shear components and the angle from the 
direction of right ascension to the lens-source direction in sky coordinates measured by the 
{\hsc} pipeline \citep{mandelbaum18, bosch+18}.
$\Sigma_{\rm crit}$ is the critical surface mass density:
\begin{equation}
\Sigma_{\rm crit} = \frac{c^2}{4\pi G} \frac{D_{\rm A}(z_{\rm s})}{D_{\rm A}(z_{\rm L}) 
D_{\rm A}(z_{\rm L}, z_{\rm s})}
\label{eq:sigmacrit}
\end{equation}
which is computed using the angular diameter distance between
the source and observer $D_{\rm A}(z_{\rm s})$, lens and observer $D_{\rm A}(z_{\rm l})$, and
source and lens $D_{\rm A}(z_{\rm L}, z_{\rm s})$. Each source galaxy is weighted by:

\begin{equation}
w_{\rm Ls} = \frac{\Sigma_{\rm crit}^{-2}}{\sigma^2_{e, {\rm Ls}} + \sigma^2_{\rm rms}}
\equiv \frac{\Sigma_{\rm crit}^{-2}}{\sigma^2_{{\rm Ls}}}
\label{eq:weight}
\end{equation}

\noindent where $\sigma_{\rm rms}$ is the intrinsic shape dispersion per component 
and $\sigma_{e, Ls}$ is the per-component shape measurement error \citep[see][]{mandelbaum+18}. 
$\mathcal{R}(R)$ is the shear responsivity 
factor\footnote{For $\Delta\Sigma$, we have verified 
that the weighting applied to $R$ should include the $\Sigma_{\rm crit}^{-2}$ 
factor as defined in Equation \ref{eq:weight}.} that describes the response of 
galaxy ellipticity to a small amount of shear:

\begin{equation}
\mathcal{R}(R) = 1 - \frac{\Sigma_{\rm Ls}^{R} w_{\rm Ls} \sigma^2_{{\rm Ls}}}{\Sigma_{\rm Ls}^{R} w_{\rm Ls}}
\label{eq:rfactor}
\end{equation}

\noindent We compute and apply $\mathcal{R}$ independently for each radial bin. 

The $[1 + K(R)]$ term is a correction for the multiplicative shear bias $m$:
\begin{equation}
K(R) = \frac{\Sigma_{\rm Ls}^{R} w_{\rm Ls} m_{\rm s}}{\Sigma_{\rm Ls}^{R} w_{\rm Ls}}
\label{eq:rfactor}
\end{equation}
where $m_{\rm s}$ is a per source value that is calibrated using simulations.
Please see \cite{mandelbaum+18} for details about the calibration of {\hsc} weak lensing catalog.

In this work, we use $10^5$ random points to compute ${\Delta\Sigma}_{\rm R}$
sampled following the {\hsc} \texttt{S16A} survey geometry. Random points are assigned redshifts
following the redshift distribution of lenses. Although
\cite{singh+17} use the boost factor ($\Sigma_{\rm Ls}^{R} w_{\rm Ls}/\Sigma_{\rm Rs}^{R} w_{\rm Rs}$)
to correct for dilution effects (see also \citealt{mandelbaum+05}),
we do not apply any boost factor corrections here. Instead, in Section \ref{subsec:zlss} 
we test whether or not the signal varies as we impose more stringent lens-source separation cuts.


We assume physical coordinates and compute $\Delta\Sigma$ in 10 logarithmically
spaced bins from 0.05\,Mpc to 15\,Mpc. Errors on  all $\Delta\Sigma$-related quantities 
are computed via bootstrap resampling. The \texttt{dsigma} code divides lenses and randoms to 
roughly equal-area regions. Here we use 40 regions with typical sizes of $\sim$ 2.5\,deg and 
compute errors with $N_{\rm Bs} = 5000$ bootstraps.

\subsection{Corrections for Photometric Redshift Bias and Dilution Factors}
\label{subsec:wlbias}

Our procedure for estimating the bias on $\Delta\Sigma$ arising from {\phz}'s
partially follows that of \citet{mandelbaum+08}, \citet{nakajima+12}, and \citet{leauthaud+17}.
We summarize our approach here.

To correct for biases in $\Delta\Sigma$ arising from {\phz} errors,
a common procedure is to use set of galaxies with spectroscopic redshifts that have been re-weighted
with appropriate color-magnitude weights (see \S\ref{subsec:prior}) 
to match the source distribution. However, as shown in Figures \ref{fig:som} and \ref{fig:sel}, the
currently available {\spz}'s in our training set are so underrepresented 
in some regions of color-magnitude space
that it is impossible to properly re-weight them ({\spz}'s) to match the source sample.

For this reason, instead of a spectroscopic redshift catalog, we build a
calibration catalog based on the set of many-band {\phz}'s
in the COSMOS field matched with observations taken at {\hsc} \texttt{Wide} depths 
with weak lensing cuts applied (see \S\ref{subsec:source}). Our assumption here is that the 
COSMOS many-band {\phz}'s
have narrow enough PDFs that they can be used to compute biases on $\Delta\Sigma$ for HSC. 
We refer to this catalog hereafter as the COSMOS calibration sample. 

We compute the bias on $\Delta\Sigma$ as follows. Let $\Delta\Sigma_{\rm P}$ ($\Sigma_{\mathrm{crit,P}}$) 
represent the value of $\Delta\Sigma$ measured with \phz's
and $\Delta\Sigma_{\rm T}$ ($\Sigma_{\mathrm{crit,T}}$) represent the true value of $\Delta\Sigma$. 
We define $f_{\rm bias} \equiv \Delta\Sigma_{\rm T}/\Delta\Sigma_{\rm P}$ and estimate it via:

\begin{equation}
f_{\rm bias} = \frac{\sum_{\rm Ls} w_{\rm Ls}
\left( \Sigma_{{\rm crit,T,Ls}}/\Sigma_{{\rm crit,P,Ls}} \right)}{\sum_{\rm Ls} w_{\rm Ls}}
\label{eq:ds_equation3}
\end{equation}

\noindent where the sum is performed over source galaxies drawn from the 
COSMOS calibration sample. Unlike other versions of this equation
(e.g. Equation A3 in \citealt{leauthaud+17}) there is no re-weighting factor to account for color mis-matches between the source sample and the calibration sample because 
our COSMOS calibration sample is already representative.

For a given lens sample, we estimate $f_{\rm bias}$ using Monte Carlo
methods by randomly drawing sources from our COSMOS calibration catalog and lens redshifts from the lens sample. 
We correct all $\Delta\Sigma$ values reported hereafter using $f_{\rm bias}$. 
This accounts for the dilution effect by sources that scatter above $z_{\rm L}$ but which are actually 
located at redshifts below $z_{\rm L}$). 

More explicitly, there are three issues: the impact of {\phz} scatter and bias for 
sources that are above the lens redshift, dilution due to sources that are below the 
lens redshift but get scattered above it due to {\phz} error, and dilution due 
to physically-associated sources. Our approach corrects for the first two of these, 
but not the third.

For our signals, typical values for $f_{\rm bias}$ are around a few percent ($\sim 2-5\%$).

\subsection{Comparing Lensing Signals}
\label{subsec:wl2samp}

One of the primary concerns in this work is the robustness of the gg lensing signal with respect to
possible {\phz} biases. Since an absolute calibration does not exist, we instead aim to 
demonstrate the robustness of the signal to various cuts and choices for lens-source separation. 
We quantify this by considering the ratio of $\Delta \Sigma$ for two different computations $i$ and $j$:

\begin{equation}
f_{i,j} \equiv \Delta \Sigma_i / \Delta \Sigma_j.
\end{equation}

This ratio test assumes that when we change how we calculate the gg lensing signals (by, e.g., tweaking
the source sample selection or the redshift estimator), the true 
$\Delta \Sigma(R)$ should be the same (i.e. $ \Delta \Sigma_i(R) = \Delta \Sigma_j(R)$ for all $R$).
This relies on the assumption that $\Delta \Sigma(R)$ does not vary much across the sample 
\textit{within the lens redshift bins}. In other words, we assume that changing the source sample 
in a way that emphasizes different redshifts within the lens sample 
does not meaningfully change $\Delta \Sigma(R)$ given the same {\phz} quality 
across both source samples.

While it is straightforward to take the ratio between two lensing signals with different source cuts, 
${\Delta \Sigma}_i$ and ${\Delta \Sigma}_j$ will be highly correlated. 
To deal with this effect, we derive the covariance matrix for $f$ via bootstrap resampling
using the same bootstrap regions as described previously.


We assume that $f_{i,j}$ is a constant (we only consider amplitude changes) 
and solve for the maximum-likelihood result (MLE). 
We fit for amplitude shifts over our full radial range (denoted $f_{\rm all}$) 
and also over the radial range $R=[0.1,1]$ Mpc (denoted $f_{\rm inner}$)
and $R=[1,10]$ Mpc (denoted $f_{\rm outer}$).



\section{Results: How Robust is the Galaxy-Galaxy Lensing Signal?}
\label{sec:gglens}

We now investigate how robust gg lensing signals are to various {\phz} estimators and
quality cuts. After exploring changes in the gg lensing sensitivity to a variety
of choices (\S\ref{subsec:zpoint}-\S\ref{subsec:zinfo}), we subsequently use the results of those choices to
define a ``fiducial''
sample (see \S\ref{subsec:fiducial}). Our results are presented in terms of
the stability of the gg lensing signal based on other possible 
choices with respect to our fiducial sample.

The various cuts that we test comprise 15 unique lens-source samples. 
These are described in detail below and summarized in Table \ref{tab:robusttests}.

Since we perform a large number of tests ($3 \times 3 \times 14 = 126$) in the
subsequent sections, many often correlated,
we might expect by pure statistical chance that some of the results reported
might have deviated from the expected null result by a large amount. 
We attempt to control against these in two ways. First, since for each lens sample we
compute $f_{\rm all}$, $f_{\rm inner}$, and $f_{\rm outer}$ under 14 configurations, 
we expect \textit{at most} one test to display outliers at $>3\sigma$ significance.
We thus adopt a $3\sigma$ threshold as reasonably indicative of a significant deviation.
In addition, we also compare the distribution of our error-normalized $f/\sigma_f$ 
values to those expected under a Gaussian distribution using a Kolmogorov-Smirnov (KS) test.
While the sample size is small, we find that for all cases our results are inconsistent with a Gaussian
distribution, with the results driven primarily by outliers.

\begin{table}
\caption{Redshift estimates and selection criteria used to construct robustness
tests for galaxy-galaxy lensing. V indicates that the quantity is 
varied for a particular test, while F indicates that it is kept fixed. These comprise a total 
of 15 unique lens-source samples.}\label{tab:robusttests}
\begin{tabular}{@{}lccccc}
\hline
Test & \S\ref{subsec:zpoint} & \S\ref{subsec:zqual}  &  \S\ref{subsec:zlss} &
\S\ref{subsec:zhighlow} & \S\ref{subsec:zinfo} \\
\hline
\hline
Photo-z Estimate \\
\hline
\hline
Mean  								& V &   &   &   &   \\
Median 								& V &   &   &   &   \\
Mode  								& V &   &   &   &   \\
Best  								& V & F & F & F & F \\
MC  								& V &   &   &   &   \\
\hline
Photo-z Quality Cut \\
\hline
\hline
\texttt{basic}  					&   & V &   &   &   \\
\texttt{medium}  					& F & V & F & F & F \\
\texttt{strict}  					&   & V &   &   &   \\
\hline
Lens-Source Separation \\
\hline
\hline
$z_{\rm{low68}}>z_{\rm lens}+0.1$  	&   &   & V &   &   \\
$z_{\rm{low68}}>z_{\rm lens}+0.2$  	&   &   & V &   &   \\
$z_{\rm{low95}}>z_{\rm lens}+0.1$  	& F & F & V & F & F \\
$z_{\rm{low95}}>z_{\rm lens}+0.2$  	&   &   & V &   &   \\
\hline
High/Low Redshift \\
\hline
\hline
All 								& F & F & F &   &   \\
\texttt{zlow} 						&   &   &   & V &   \\
\texttt{zhigh}  					&   &   &   & V & F \\
\hline
{\phz} Origin \\
\hline
\hline
All 								& F & F & F & F &   \\
\texttt{plow} 						&   &   &   &   & V \\
\texttt{pmed} 						&   &   &   &   & V \\
\texttt{phigh} 						&   &   &   &   & V \\
\end{tabular}
\end{table}

\subsection{Lens Sample}
\label{subsec:lens}

We use all galaxies with spectroscopic redshifts (both the LOW-Z and CMASS samples) from the 
Sloan Digital Sky Survey (SDSS) II and III Baryon Acoustic Oscillation Survey (BOSS) \citep{abazajian+09,
eisenstein+05} that overlap with the {\hsc} survey footprint.
We apply the same geometric masks to the lens sample 
that were used when constructing the source sample (see \S\ref{subsec:source}).

\begin{figure*}
\begin{center}
\includegraphics[width=\textwidth]{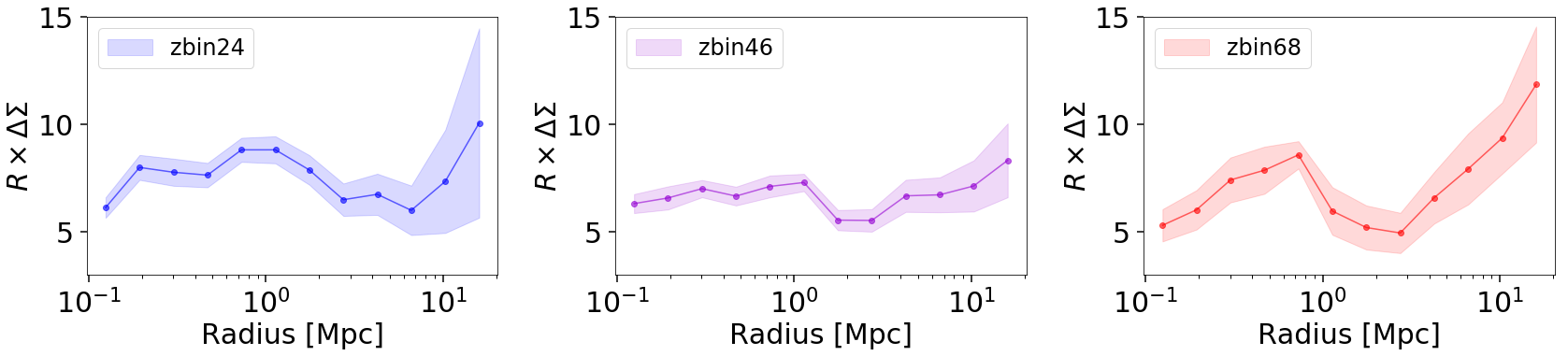}
\end{center}
\caption{$\Delta \Sigma(R)$ signals computed for our fiducial sample (defined in \S\ref{subsec:fiducial})
from BOSS CMASS and LOWZ lenses from $0.2 < z_{\rm lens} \leq 0.4$ (left, blue), $0.4 < z_{\rm lens} \leq 0.6$
(middle, purple), and $0.6 < z_{\rm lens} \leq 0.8$ (right, red). The mean values are highlighted as solid lines and
points, while the shaded region encompasses the 1-$\sigma$ errors.
}\label{fig:ds}
\end{figure*}

To explore the stability of the lensing signal as a function of redshift, we group the lens population
into three separate redshift bins:
\begin{itemize}
\item \texttt{zbin24}: $0.2 < z < 0.4$
\item \texttt{zbin46}: $0.4 < z < 0.6$
\item \texttt{zbin68}: $0.6 < z < 0.8$
\end{itemize}
They contain $\approx$ 4000, 12000, and 4000 lenses, respectively.
All the tests described below and summarized in Table \ref{tab:robusttests} 
were computed for each redshift bin, leading to a total of 45 $\Delta \Sigma(R)$
measurements. The gg lensing signal for our fiducial sample 
(see \S\ref{subsec:fiducial}) in each redshift bin is shown in Figure \ref{fig:ds}.

\subsection{Photometric Redshift Point Estimates}
\label{subsec:zpoint}

In lensing analyses, $\Delta \Sigma$ is often computed with respect to fixed point estimates
derived from the {\phz} PDFs to avoid having to integrate over all {\phz} PDFs $P(z|g)$'s.

We study whether or not the particular choice of a point source estimate impacts the gg lensing signal.
We compare five point estimates in this paper:
\begin{itemize}
\item $z_{\rm mean}$, the first moment (mean) of the {\phz} PDF,
\item $z_{\rm med}$, the 50th percentile (median) of the {\phz} PDF,
\item $z_{\rm mode}$, the redshift corresponding to the maximum value of the {\phz} PDF,
\item $z_{\rm best}$, the redshift estimator that minimizes 
the loss assuming a Lorentzian kernel in
$\Delta z / (1+z)$ with a width of $\sigma = 0.15$ 
(see \citealt{tanaka+18} for additional details), and
\item $z_{\rm mc}$, a Monte Carlo draw from the {\phz} PDF.
\end{itemize}
A comparison with integrating over the PDF is beyond the scope of this work and is discussed
further in More et al. (in prep.).

Most of these point estimates have been used to varying degrees in weak lensing analyses 
in the literature, each with various benefits and drawbacks. Here we will briefly outline the
arguments for each estimator (see also \citealt{tanaka+18}).

While beyond the scope of this paper, it is well known that the mean
estimate $z_{\rm mean}$ is the optimal point estimate for a PDF assuming ``squared error'' ($L_2$) loss. In
other words, if we introduce a penalty proportional to $(z_{\rm est} - z_{\rm true})^2$ and assume
$z_{\rm true}$ follows our PDF, then $z_{\rm est} = z_{\rm mean}$ is the estimator that is ``best''
given the PDF. This particular result is optimal for Gaussian distributions.

In general, however, most {\phz} PDFs are not Gaussian, but instead can have asymmetric tails and/or
extended shapes. The mean $z_{\rm mean}$ is particularly sensitive to these tails, and so estimates that
are more ``robust'' are sometimes preferred. As with the mean, it can likewise be shown that the
median $z_{\rm med}$ is the optimal point estimate under ``absolute'' ($L_1$) loss where the penalty
is proportional to $|z_{\rm est} - z_{\rm true}|$. This reduced penalty makes $z_{\rm med}$
less sensitive to the tails. The mode $z_{\rm mode}$ can likewise be shown to be
the optimal point estimate under ``unforgiving'' 
loss ($L_0$) where the penalty is maximized and constant for all $z_{\rm est} \neq z_{\rm true}$. This
penalty makes $z_{\rm mode}$ only sensitive to the peak of the PDF where the probability is maximized.

While these various estimators are optimal under different assumptions for how much we want to penalize
``incorrect'' guesses, none of them are specifically tuned for {\phz} estimation. In particular, most
PDFs and {\phz} applications tend to have a dependence on $|z_{\rm est} - z_{\rm true}|/(1+z_{\rm true})$
rather than just $|z_{\rm est} - z_{\rm true}|$, and also care about being accurate relative
to a given ``tolerance'' $\sigma$. $z_{\rm best}$ is the point estimate that minimizes the loss
relative to these conditions.

Finally, we may want a point estimate that ``explores'' the entire PDF, rather than attempting to
``summarize'' it. Assuming ``uniform'' loss (i.e. a flat penalty everywhere), any
Monte Carlo sample $z_{\rm mc}$ from the PDF serves as a reasonable point estimate. These may better capture
the behavior of PDFs by allowing us to probe, e.g., the tails of the distribution but lead to some
(additional) amount of random noise being introduced.

\begin{figure*}
\begin{center}
\includegraphics[width=\textwidth]{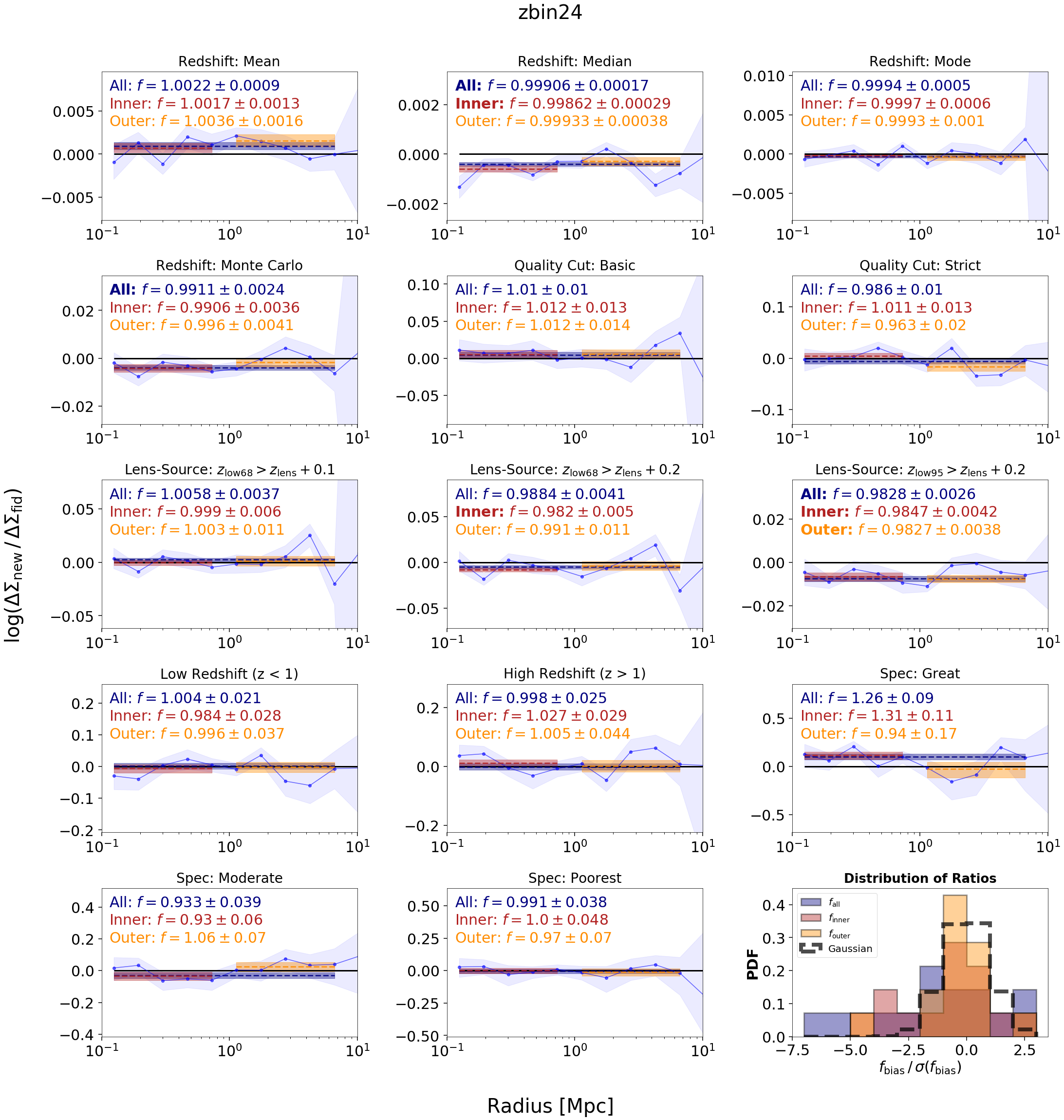}
\end{center}
\caption{Bootstrap ratios $f_{\rm new,\, fid}$ between new $\Delta \Sigma_{\rm new}(R)$ 
signals computed using the \texttt{zbin24} lens sample ($0.2 < z_{\rm lens} \leq 0.4$) 
relative to our fiducial sample $\Delta \Sigma_{\rm fid}(R)$. 
The null result ($f=1$) is shown as a solid black line, while
the maximum-likelihood estimator for $f_{\rm all}$ (dark blue) and
$f_{\rm inner}$ and $f_{\rm outer}$ (dark red) for each sample are shown as dotted lines 
and listed in the upper-left-hand corner of each plot. 
1-$\sigma$ errors on all quantities are shown as shaded regions. A 1-D histogram of the error-normalized
distribution of $f$, $f_{\rm inner}$, and $f_{\rm outer}$ is displayed 
in the bottom-right corner along with a Gaussian distribution for reference.
Most of the computed ratios are consistent with the expected null result at 3$\sigma$; those that
disagree are highlighted in bold. 
Although these disagreements are statistically significant, some
(e.g., with respect to $z_{\rm med}$) are negligible in practice since their impact is
$\lesssim 1\%$. In general, $f_{\rm outer}$ is
more unbiased for all samples than $f$ and $f_{\rm inner}$.
See \S\ref{sec:gglens} for additional discussion and details.
}\label{fig:dds1}
\end{figure*}

\begin{figure*}
\begin{center}
\includegraphics[width=\textwidth]{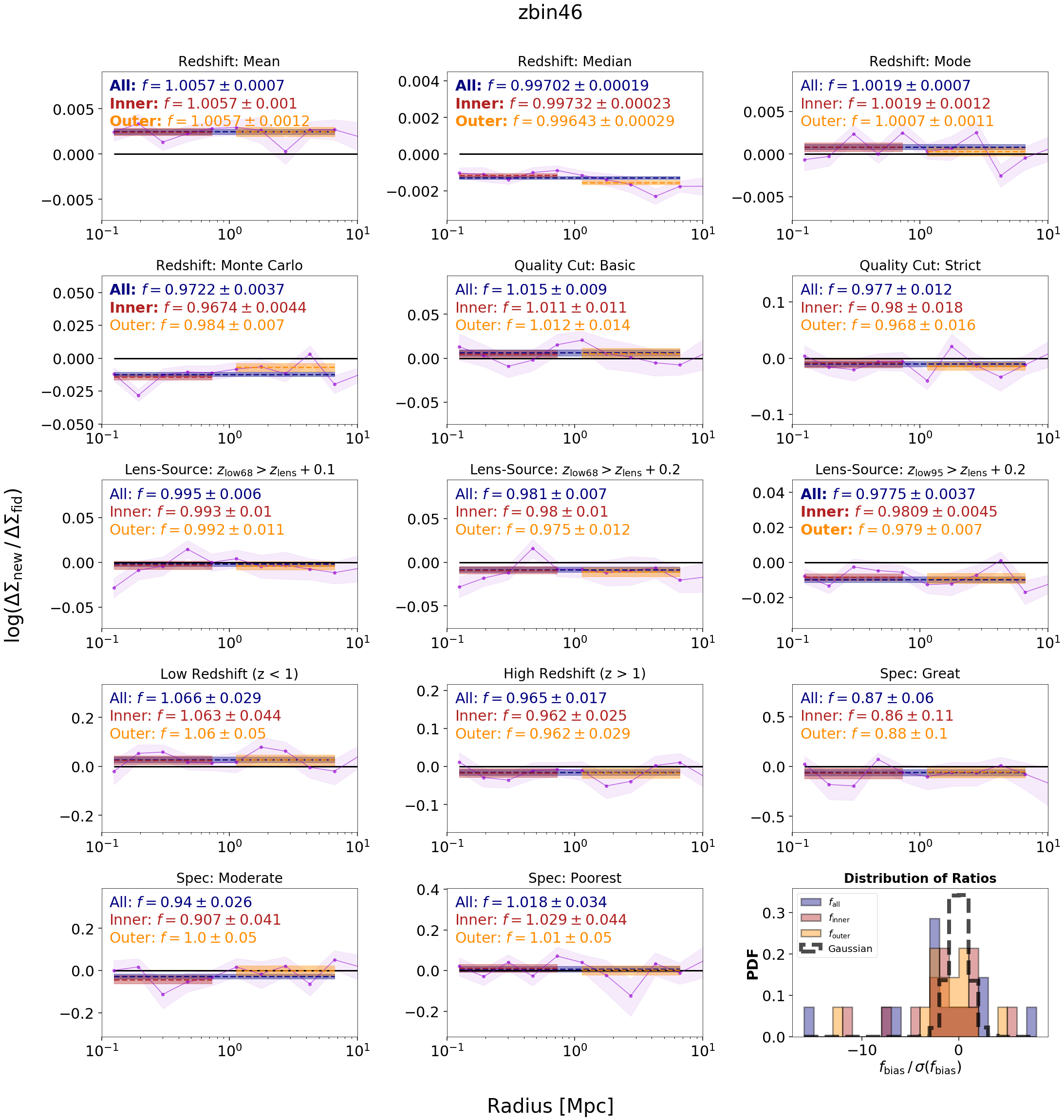}
\end{center}
\caption{As Figure \ref{fig:dds1}, but for \texttt{zbin46} ($0.4 < z_{\rm lens} \leq 0.6$).
}\label{fig:dds2}
\end{figure*}

\begin{figure*}
\begin{center}
\includegraphics[width=\textwidth]{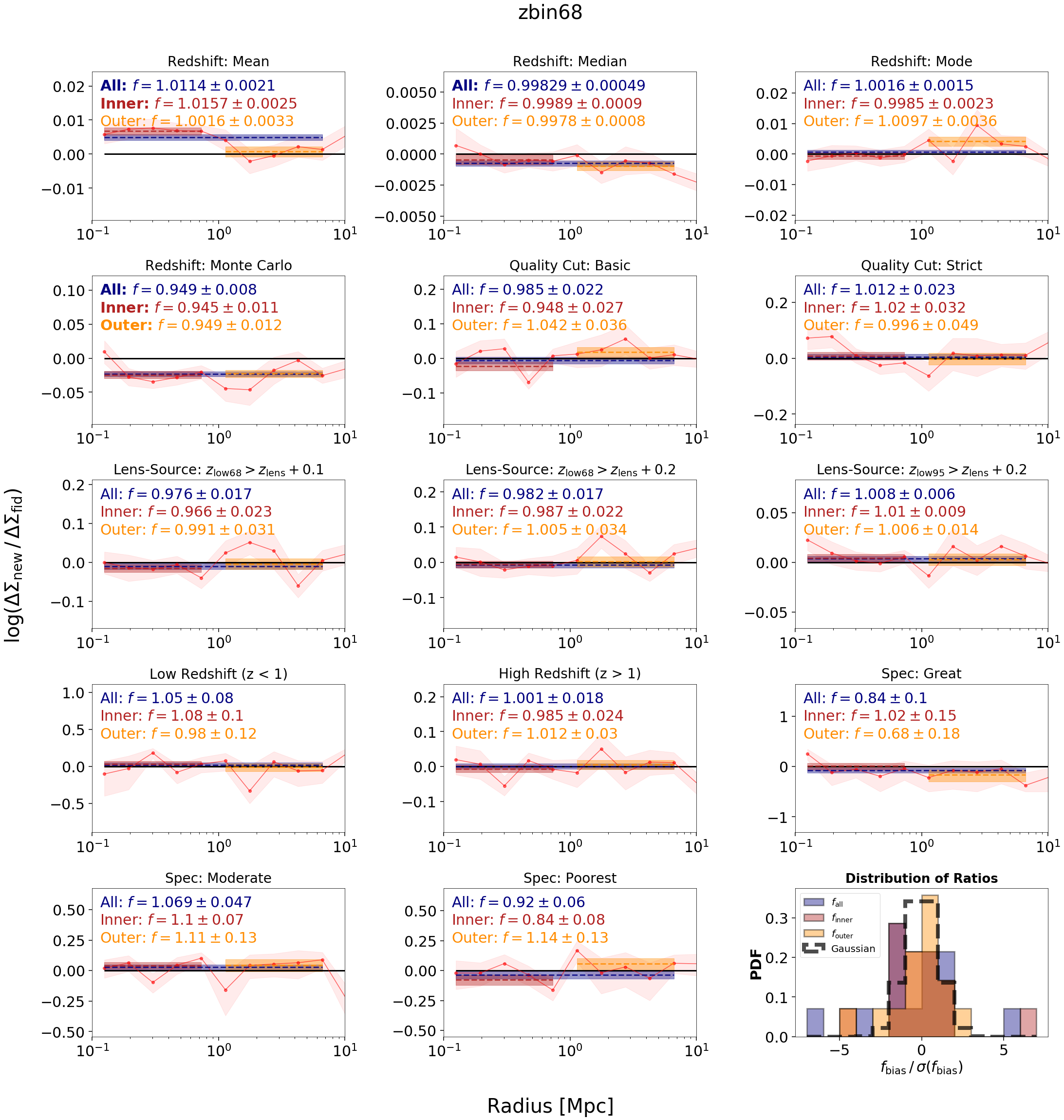}
\end{center}
\caption{As Figure \ref{fig:dds1}, but for \texttt{zbin68} ($0.6 < z_{\rm lens} \leq 0.8$).
}\label{fig:dds3}
\end{figure*}

The $\Delta\Sigma$ ratio estimates computed based on
each of these various redshift point estimates with respect to our
fiducial sample in each redshift bin
are shown in the top two rows of Figures \ref{fig:dds1}, \ref{fig:dds2}, and \ref{fig:dds3}.
We find that, with the exception of $z_{\rm mc}$,
all of these choices result in negligible ($\lesssim 1\%$), albeit
sometimes statistically significant (at 3$\sigma$), differences in the computed $\Delta \Sigma$.
This is likely due to the general quality of our PDFs,
which are reasonably well-constrained and unimodal for the majority of objects (see Figure
\ref{fig:pz1}) and also well-calibrated against the expected underlying redshift distribution
(Figure \ref{fig:pz3}), leading to very similar point estimates.

In general, using Monte Carlo redshifts $z_{\rm mc}$ tends to lead to an underestimate of
the $\Delta \Sigma$ signal by an increasing amount as a function of the lens redshift. This
is due to the (exponentially) increasing sensitivity of the $\Delta\Sigma$ signal at close
lens-source separations as well as increasing {\phz} uncertainties at higher redshifts.
Since Monte Carlo redshifts scatter sources around based on their PDFs, these
tend to dilute the computed signals relative to the more ``stable'' point estimates above.

Since there is no (relevant) statistical difference between the {\phz} point estimates excluding $z_{\rm mc}$,
we decide to use the $z_{\rm best}$ estimate due to its superior performance 
relative to the other estimators when predicting redshifts for individual objects
within the full {\hsc} \texttt{S16A} \texttt{Wide} sample. For additional comparisons between
the per-object accuracy of these {\phz} point estimates, see \citet{tanaka+18}.

\subsection{Photometric Redshift Uncertainties}
\label{subsec:zqual}

Although our {\phz} PDFs are well-calibrated with respect to the underlying
redshift distribution (\S\ref{subsec:cv}), in general using point estimates to summarize broad PDFs can
lead to distortions in the underlying redshift population \citep{carrascokindbrunner14c}. In addition,
broad {\phz} PDFs where the probability density spans a large redshift range are generally seen as more
unreliable, with more potential for miscalibrations 
that can lead to under/overestimated uncertainties on
the prediction compared to narrower PDFs. As a result, it is common in many gg lensing analyses to
remove ``unreliable'' {\phz}'s based on the width of their PDFs.

We define two main sources of uncertainty that contribute to unreliable PDFs.
The first is \textit{systemic} uncertainty: having a poor understanding of the object
in question and therefore an unreliable redshift prediction. This can occur if the
object is not well-represented within the training set, which leads to them
having large $\chi^2$ values when comparing to their closest color-magnitude neighbors.
We can exploit this fact to flag and remove these sources explicitly.

The second source of uncertainty is \textit{statistical} uncertainty: utilizing
a point estimate that does not accurately represent the PDF. This can occur if the
redshift PDF is overly broad or multi-modal with several possible redshift solutions.
We quantify this source of uncertainty by defining the ``risk'' \citep{tanaka+18} that
the point estimate is incorrect is the integral over the PDF with respect to
the associated loss
\begin{equation}
    R(z_{\rm phot}) = \int P(z) L(z, z_{\rm phot}) \, dz
\end{equation}
where the particular loss function
\begin{equation}
    L(z, z_{\rm phot}) = 1 - \frac{1}{1+\left(\frac{z_{\rm phot} -z}{\gamma(1+z)}\right)^2}
\end{equation}
is taken to be a Lorentzian kernel with a width of $\gamma=0.15$. The
best redshift estimate and the associated risk $z_{\rm risk}$ are then defined jointly as
\begin{equation}
    z_{\rm risk} = \min\{R(z)\} = R(z_{\rm best}
\end{equation}
Sources with higher $z_{\rm risk}$ generally have broader PDFs with multiple peaks.
See \citet{tanaka+18} for additional discussion.

We divide our sample into a number of sub-samples based on a range of {\phz} quality cuts. These are:
\begin{itemize}
\item \texttt{basic}: $\chi^2_5 \leq 6$. This is the $\chi^2_5$ value corresponding to the 95\% cut
discussed in \S\ref{subsec:pznew} since $P(\chi^2_5 \leq 6) \approx 0.95$ for a chi-square distributed
random variable with 5 degrees of freedom. As expected, this removes $\sim 5\%$ of sources. The majority
of these sources are at brighter magnitudes and lower redshifts and do not not contribute significantly
to the gg lensing signal.
\item \texttt{medium}: In addition to \texttt{basic}, this selection also imposes a cut on the ``risk''
of a particular {\phz} point estimate $z_{\rm risk} < 0.25$.
This generally removes overly broad PDFs and leaves $\sim 75\%$ of the sample.
\item \texttt{strict}: In addition to \texttt{basic}, this selection imposes a stricter
cut of $z_{\rm risk} < 0.15$, restricting our estimates to even narrower PDFs than \texttt{medium}.
This leaves $\sim 60\%$ of the sample.
\end{itemize}

The $\Delta\Sigma$ ratio estimates computed for each of these global {\phz} quality cuts
with respect to our fiducial sample are shown in the second row of Figures \ref{fig:dds1}, 
\ref{fig:dds2}, and \ref{fig:dds3}. We find that the computed $\Delta \Sigma$ signals appear insensitive
to the global {\phz} quality cut chosen, and are statistically consistent with the null result (at 3$\sigma$). 
As with \S\ref{subsec:zpoint}, this is likely due to the fact that our PDFs are both
relatively well-constrained and well-calibrated for the majority of our sample, especially since any
outlying PDFs are removed by our initial \texttt{basic} quality cuts.

Since our performance is similar across different global {\phz} quality cuts, we opt to use our
\texttt{medium} cut for our fiducial sample to compromise between sample size and PDF quality.

\subsection{Lens-Source Separation}
\label{subsec:zlss}

In addition to possible biases based on how the {\phz} point estimates
trace the underlying source population,
it is also imperative to ensure that any possible differences we observe 
are not dominated by dilution effects
or contamination from correlated objects (see \S\ref{subsec:wlbias}). 
This is often done by imposing cuts
that aim to ensure that the bulk of any source galaxy PDF lies behind the lens population
(e.g., \citealt{medezinski+18}; see also \S\ref{subsec:zpoint}).

We parameterize this cut using two parameters. 
The first is a summary statistic detailing the redshift below which
X\% ($z_{\rm{lowX}}$) of the source galaxy PDF lies, which we use to
establish how confidently we can place a source galaxy behind a given lens. In other words,
X\% of the {\phz} PDF is above a redshift of $z_{\rm{lowX}}$, which should be greater than
the redshift of the lens $z_{\rm lens}$.
The second is a ``buffer'' $\Delta z_{\rm lens}$
to establish a minimum separation threshold between the source $z_{\rm lowX}$
and the lens $z_{\rm lens}$. This term is used to avoid
being extremely sensitive to {\phz} biases and possible miscalibrations in the corresponding PDFs
since the dependence of $\Sigma_{\rm crit}$ is highly non-linear
when $z_{\rm source}$ and $z_{\rm lens}$ are very close together.

We test four lens-source separation cuts:
\begin{itemize}
\item $z_{\rm{low68}} > z_{\rm lens} + 0.1$
\item $z_{\rm{low68}} > z_{\rm lens} + 0.2$
\item $z_{\rm{low95}} > z_{\rm lens} + 0.1$
\item $z_{\rm{low95}} > z_{\rm lens} + 0.2$
\end{itemize}
for $X=68\%$ and 95\% (roughly 1 and 2-sigma) and $\Delta z_{\rm lens}=0.1$ and 0.2, 
listed roughly in order from most aggressive to most conservative. 

Our results are shown in the third and fourth rows of Figures \ref{fig:dds1},
\ref{fig:dds2}, and \ref{fig:dds3}. We find that all cases are 
consistent (at 3$\sigma$) with the null result with the exception 
of the $z_{\rm{low95}}+0.2 > z_{\rm lens}$
cut for the \texttt{zbin24} lens sample, which is smaller by $\approx 1.5\%$.
As the most conservative cut in the lowest redshift bin (which should be least-sensitive
to {\phz} issues), it is somewhat surprising that we see a noticeable suppression.
While it is possible that this is just statistical noise, it might also be the case that
the {\phz}'s in the training set have systematic discrepancies
at intermediate redshifts that are accentuated when only the low-$z$ sources are removed.

In general, however, these results support
the gg lensing signal being mostly insensitive to these 
specific combination of lens-source separation cuts
for our {\hsc} \texttt{S16A} data. 
This implies that the majority of our sources are correctly selected to be
behind the bulk of the lens sample, providing additional (indirect) support that our {\phz} PDFs 
are well-calibrated. These results also suggest that our gg lensing signals 
do not require any boost factor corrections.

Given that these lens-source separation cuts perform comparably, we again opt to use a compromise for
our fiducial sample by choosing $X=95\%$ and $\Delta z_{\rm lens}=0.1$.



\subsection{High and Low Redshift Sources}
\label{subsec:zhighlow}

One additional concern is our gg lensing analysis may be sensitive to degrading {\phz} quality
as a function of redshift. This is in general due to a combination of spectroscopic incompleteness
at higher redshifts and fainter magnitudes as well as broader PDFs arising from noisier photometry
(see \S\ref{sec:som}).
This is a particularly acute concern for this work due to our reliance on many-band {\phz}'s
at the magnitudes and redshifts probed by a significant majority of our weak lensing source galaxies.

To investigate this effect, we divide our source sample into high-redshift and low-redshift
samples to investigate this effect defined by:
\begin{itemize}
\item \texttt{zlow}: $z_{\rm best} \leq 1$, which leaves $\sim 50\%$ of the sample.
\item \texttt{zhigh}: $z_{\rm best} > 1$, which leaves $\sim 50\%$ of the sample.
\end{itemize}

The results for our high and low-redshift samples are shown in the fourth row of
Figures \ref{fig:dds1}, \ref{fig:dds2}, and \ref{fig:dds3}. Although we find the $\Delta\Sigma$
signals from \texttt{zlow} to be systematically higher than those from \texttt{zhigh}, the
effect is not statistically significant and both agree with the null result (at 3$\sigma$).
This provides us with confidence that
we can utilize {\phz}'s for source galaxies at all redshifts when constructing our fiducial sample.

\subsection{Origin of Training Redshifts}
\label{subsec:zinfo}

One benefit of the KMCkNN framework outlined in
\S\ref{sec:photoz} is that we actually have a \textit{direct} proxy of spectroscopic incompleteness through
metrics such as \texttt{Fphot} and \texttt{Pphot}, in addition to indirect proxies such as
the high/low redshift split used in \S\ref{subsec:zhighlow}. This allows us to examine how robust our
gg lensing signals are depending on the \textit{information content} used to estimate the
{\phz}'s of individual source galaxies.

One complication of using \texttt{Pphot} to select source galaxies directly is that it tends to be
strongly correlated with magnitude and redshift, with sources with lower redshifts
and brighter magnitudes tending to also have lower \texttt{Pphot}. We attempt to alleviate this
issue by limiting our analysis to the \texttt{zhigh} subset of galaxies ($z_{\rm best} > 1$). While
this substantially reduces the sample (by 50\%), it mitigates
some of the extreme differences that can arise due to these effects.

As a compromise between preserving number density and maximizing differences between sources that
are many-band {\phz}-dominated versus those that are not, we ultimately split our source galaxies
into three subsamples based on {\spz} and {\gprz} information content:
\begin{itemize}
\item ``Great'': $\texttt{Pphot} < 0.5$ 
(i.e. $> 50\%$ of information comes from {\spz}'s and {\gprz}'s),
which leaves $\sim 10\%$ of the sample.
\item ``Moderate'': $0.5 \leq \texttt{Pphot} < 0.85$ (moderately {\phz}-dominated),
which leaves $\sim 15\%$ of the sample.
\item ``Poorest'': $\texttt{Pphot} \geq 0.85$ (completely {\phz}-dominated),
which leaves $\sim 25\%$ of the sample.
\end{itemize}

The results for these subsamples are shown in the bottom two rows
of Figures \ref{fig:dds1}, \ref{fig:dds2}, and \ref{fig:dds3}. 
We find these signals are entirely consistent with the null result (at 3$\sigma$).
\textit{This demonstrates that our gg lensing signals are 
stable to the origin of the training redshifts
(e.g. spec-z, photo-z, etc.) used to compute redshifts for source galaxies}.
We note, however, that the {\spz} samples used to train our {\phz}'s
tend to have targeted very specific populations of galaxies even at 
higher redshift compared with the broader photometric sample (see \S\ref{sec:som}).
This can lead to low \texttt{Pphot} serving as a
proxy for selecting galaxy samples in particular regions of
color-magnitude space (and thus having different intrinsic properties). These 
changes in the underlying galaxy population could mask some of the expected impacts from 
many-band {\phz}'s alone and make it difficult to extrapolate conclusions beyond this current work.

In general, as we find that the computed $\Delta\Sigma$ signals
are consistent with null results across all lens redshift and \texttt{Pphot} subsamples,
we opt to include all sources when constructing our fiducial catalog.

\subsection{Fiducial Lensing Cuts}
\label{subsec:fiducial}

Based on the results above, 
we now define the fiducial sample that all other samples are compared
to in Figures \ref{fig:dds1}, \ref{fig:dds2}, and \ref{fig:dds3}.
Our reasoning is as follows:
\begin{itemize}
\item All point estimates (excluding $z_{\rm mc}$) investigated in this work give gg lensing signals
with similar amplitudes. We thus opt to use the $z_{\rm best}$ point estimates (\S\ref{subsec:zpoint})
given their improved performance across the broader photometric sample as outlined in \citet{tanaka+18}.
\item Since the three basic quality cuts give similar $\Delta \Sigma$ estimates, we select the
\texttt{medium} {\phz} quality cuts (\S\ref{subsec:zqual}) as a fiducial choice. This represents a
compromise between retaining a larger sample size and removing overly broad {\phz} PDFs.
\item All four combinations of lens-source separation cuts give gg lensing signals that are consistent
with each other. As with the global {\phz} cuts, we decide to then compromise by selecting
$z_{\rm low95}+0.1 > z_{\rm lens}$ (\S\ref{subsec:zlss}), which guarantees the vast majority of the
PDF is located behind the lens while being slightly less conservative 
about the enforced $\Delta z_{\rm lens}$
separation.
\item Our tests over high ($z_{\rm best} \geq 1$) and low ($z_{\rm best} < 1$) 
subsamples of source galaxies do not show
any sign of distortion by {\phz} biases arising from changing populations of objects in
our training data (\S\ref{subsec:zhighlow}). To maximize sample size, we thus opt to use 
galaxies at all available redshifts.
\item Finally, our tests using subsamples binned by \texttt{Pphot} 
at $z_{\rm best} > 1$ also do not find evidence for differences
in $\Delta \Sigma$ among the varying subsamples (\S\ref{subsec:zinfo}).
As a result, we opt to include all {\phz}'s regardless of their spectroscopic information content.
\end{itemize}

These cuts are implemented as defaults in \texttt{dsigma}.

\section{Conclusion}
\label{sec:conc}

Determining accurate photometric redshifts (\phz's) remains a key challenge for deep lensing surveys
such as {\hsc} and LSST. At the depths probed by {\hsc}, 
there remains a dearth of spectroscopic redshifts
available for training, validating, and testing {\phz} methods across the colors and magnitudes covered
by weak lensing photometric samples. To reach the required coverage to compute {\phz}'s to
these objects, the {\hsc} {\phz} team constructed 
a heterogeneous training set derived from an amalgamation
of public {\spz} and {\gprz} surveys along with \phz's derived from deep, many band COSMOS data.

Since mixing \spz's and high-quality alternatives (\gprz's, \phz's) will likely occur in 
future surveys, in this paper we sought to thoroughly investigate their impact on gg lensing analyses through
a variety of methods. Our conclusions are as follows:
\begin{enumerate}
\item Using Self-Organizing Maps (SOMs), we examine the color/magnitude-space 
coverage of our {\hsc} training data relative to the {\hsc} \texttt{S16A} weak lensing photometric sample
(\S\ref{sec:som}).
We find that, as expected, our {\spz} coverage is highly non-representative relative to the overall sample,
with the majority of our redshift information for ``typical'' galaxies in the weak lensing photometric
sample coming from many-band {\phz}'s.
\item We then investigated whether current spectroscopic survey strategies, which seek to systematically
fill in underpopulated regions of color space, can resolve this problem (\S\ref{subsec:magcolor}).
We find that the assumption
that the intrinsic redshift distribution at fixed color is constant as a function of magnitude does
not always hold. This mismatch implies that certain regions of color space will likely require {\spz}'s that
also probe the magnitude distribution of future weak lensing samples, complicating current efforts. This effect
is in addition to redshift-dependent success rates at fixed color and magnitude.
\item Based on these results, in \S\ref{sec:photoz} we develop a hybrid machine learning/Bayesian
framework for tracking how subsets of
galaxies in our training set contribute to individual {\phz} predictions explicitly as a function of
magnitude. We show that in \S\ref{sec:pzchecks} our approach gives reasonable {\phz} predictions and
well-calibrated PDFs.
\item Using our fits, we are able to define metrics to reject objects that are poorly represented by
the training data and further identify how ``reliable'' results are based on how significantly
many-band {\phz}'s contribute to the derived PDFs (\S\ref{subsec:cv}). The results imply that {\phz}'s
computed for objects with $i \lesssim 23$ tend to be {\spz}-dominated, while those at $i \gtrsim 23$ tend
to be {\phz}-dominated.
\item Finally, using the full sample of LOWZ and CMASS BOSS galaxies, we investigate the impact various
{\phz} estimators, quality cuts, lens-source separation constraints, redshift subsamples, and
spectroscopic information content can have on gg lensing signals. We find that most cases give
results that are consistent with a fiducial baseline sample, indicating that biases in the
gg lensing signal due to {\phz} bias and scatter are sub-dominant
to statistical uncertainties in the {\hsc} \texttt{S16A} weak lensing data.
\end{enumerate}

Although we do not find cause for concern in the analysis presented here, we hope that these methods
can be used in future work to investigate similar issues when dealing with larger, deeper, and more complex
samples leading up to future precision cosmology-oriented surveys such as LSST.

\section*{Acknowledgments}

The authors would like to thank the referee for feedback that helped to improve
the quality of this work.
JSS is eternally grateful to Rebecca Bleich for her patience, assistance, and support. 
JSS thanks Charlie Conroy, Doug Finkbeiner, Lars Hernquist, Jean Coupon,
and Mara Salvato for helpful feedback that improved the quality of this work.
JSS acknowledges financial support from the CREST program, which is funded by the 
Japan Science and Technology (JST) Agency, for partially supporting this work, as well as the UC-Santa
Cruz Astronomy and Astrophysics Department and graduate student body for their kindness and hospitality.

This material is based upon work supported by the National Science 
Foundation under Grant No. 1714610. 
JSS is supported by the National Science Foundation Graduate Research
Fellowship Program. 
AL acknowledges support from the David and Lucille Packard foundation, 
and from the Alfred P. Sloan foundation.
RM is supported by the Department of Energy Cosmic Frontier program, grant DE-SC0010118.
A portion of this research was carried out at the Jet Propulsion Laboratory, 
California Institute of Technology, 
under a contract with the National Aeronautics and Space Administration.

The Hyper Suprime-Cam (HSC) collaboration includes the astronomical
communities of Japan and Taiwan, and Princeton University. The HSC instrumentation
and software were developed by the National Astronomical
Observatory of Japan (NAOJ), the Kavli Institute for the Physics and
Mathematics of the Universe (Kavli IPMU), the University of Tokyo, the
High Energy Accelerator Research Organization (KEK), the Academia
Sinica Institute for Astronomy and Astrophysics in Taiwan (ASIAA), and
Princeton University. Funding was contributed by the FIRST program
from Japanese Cabinet Office, the Ministry of Education, Culture, Sports,
Science and Technology (MEXT), the Japan Society for the Promotion of
Science (JSPS), Japan Science and Technology Agency (JST), the Toray
Science Foundation, NAOJ, Kavli IPMU, KEK, ASIAA, and Princeton
University.
  
Based on data collected at the Subaru Telescope and retrieved from
the HSC data archive system, which is operated by Subaru Telescope and
Astronomy Data Center, National Astronomical Observatory of Japan.

The authors wish to recognize and acknowledge the very significant cultural role
and reverence that the summit of Maunakea has always
had within the indigenous Hawaiian community. We are
most fortunate to have the opportunity to use data collected from observations
taken from this mountain.



\bibliography{photoz}


\bsp	
\label{lastpage}
\end{document}